 \documentclass[12pt]{article}
  \usepackage{epsfig}
   \usepackage {graphicx}
 \usepackage[latin1]{inputenc}
 \usepackage{amsfonts}
 \usepackage{amssymb}
 \usepackage{amsthm}
\newcommand{\inc}{{\it i}}

 \newcommand{\be}{\begin{equation}}
 \newcommand{\ee}{\end{equation}}
 \newcommand{\ba}{\begin{eqnarray}}
 \newcommand{\ea}{\end{eqnarray}}
 \newcommand{\efbold}{\mbox{{\boldmath $\vec f$}}}
 
 \newcommand{\erbold}{\mbox{{\boldmath $\vec r$}}}
 \newcommand{\Phibold}{\mbox{{\boldmath $\vec \Phi$}}}
 \newcommand{\mubold}{\mbox{{\boldmath $\vec \mu$}}}
 \newcommand{\omegabold}{\mbox{{\boldmath $\vec \omega$}}}
 \newcommand{\taubold}{\mbox{{\boldmath $\vec \tau$}}}

 \newcommand{\doterbold}{\dot{\textbf {\mbox{\boldmath $\vec{\boldmath r}$}} }}

\begin{document}
\title{
%
 {\Large{\textbf{
  The theory of canonical perturbations
  applied to attitude~dynamics~and~to~the~Earth~rotation.\\
  Osculating~and~nonosculating~Andoyer~variables.
 }}}\\ }
 \author{
 {\Large{Michael Efroimsky}}\\ {\small{US Naval Observatory,
 Washington DC 20392 USA
 }}\\
and\\
{\Large{Alberto Escapa}}\\ {\small{Departamento de Matem{\'a}tica Aplicada, Universidad
de Alicante,
 Alicante E-03080 Spain}}\\
  }
  \maketitle
 \begin{abstract}

 In the method of variation of parameters we express the Cartesian
 coordinates or the Euler angles as functions of the time and six
 constants. If, under disturbance, we endow the ``constants" with time
 dependence, the perturbed orbital or angular velocity will consist of a
 partial time derivative and a convective term that includes time
 derivatives of the ``constants". The Lagrange constraint, often imposed
 for convenience, nullifies the convective term and thereby guarantees
 that the functional dependence of the velocity on the time and
 ``constants" stays unaltered under disturbance. ``Constants" satisfying
 this constraint are called osculating elements. Otherwise, they are
 simply termed orbital or rotational elements. When the equations for the
 elements are required to be canonical, it is normally the Delaunay
 variables that are chosen to be the orbital elements, and it is the
 Andoyer variables that are typically chosen to play the role of
 rotational elements. (Since some of the Andoyer elements are
 time-dependent even in the unperturbed setting, the role of ``constants"
 is actually played by their initial values.) The Delaunay and Andoyer
 sets of variables share a subtle peculiarity: under certain circumstances
 the standard equations render the elements nonosculating.

 In the theory of orbits, the planetary equations yield nonosculating
 elements when perturbations depend on velocities.
 To keep the elements osculating, the equations must be amended with extra
 terms that are {\textbf{not}} parts of the disturbing function
 (Efroimsky \& Goldreich 2003, 2004; Efroimsky 2005, 2006). It
 complicates both the Lagrange- and Delaunay-type planetary
 equations and makes the Delaunay equations noncanonical.

 In attitude dynamics, whenever a perturbation depends upon the angular velocity
 (like a switch to a non-inertial frame), a mere amendment of the Hamiltonian
 makes the equations yield nonosculating Andoyer elements. To make them
 osculating, extra terms should be added to the equations (but then the
 equations will no longer be canonical).

Calculations in nonosculating variables are mathematically valid, but their
physical interpretation is not easy. Nonosculating orbital elements
parameterise instantaneous conics \textbf{not} tangent to the orbit. (A
nonosculating {\it i} may differ much from the real inclination of the orbit,
given by the osculating {\inc}$\,$.) Nonosculating Andoyer elements correctly
describe perturbed attitude, but their interconnection with the angular
velocity is a nontrivial issue.
The Kinoshita-Souchay theory tacitly employs nonosculating Andoyer elements. For
this reason, even though the elements are introduced in a precessing frame, they
nevertheless return the inertial velocity, not the velocity relative to the
precessing frame. To amend the Kinoshita-Souchay theory, we derive the
precessing-frame-related directional angles of the angular velocity relative to the
precessing frame.

The loss of osculation should not necessarily be considered a flaw of the
Kinoshita-Souchay theory, because in some situations it is the inertial, not the
relative, angular velocity that is measurable (Schreiber et al. 2004, Petrov 2007).
Under these circumstances, Kinoshita's formulae for the angular velocity should be
employed (as long as they are rightly identified as the formulae for the
{\emph{inertial}} angular velocity).

 %

\end{abstract}

 \section{The Hamiltonian approach to rotational dynamics}

 \subsection{Historical preliminaries}

The perturbed rotation of a rigid body has long been among the key topics of
both spacecraft engineering (Giacaglia \& Jefferys 1971; Zanardi \& Vilhena de
Moraes 1999) and planetary astronomy (Kinoshita 1977; Laskar \& Robutel 1993;
Touma \& Wisdom 1994; Mysen 2004, 2006). While free spin (the Euler-Poinsot
problem) permits an analytic solution in terms of the elliptic Jacobi
functions, perturbed motion typically requires numerical treatment, though
sometimes it can be dealt with by analytical means (like, for example, in
Kinoshita 1977$\,$). Perturbation may come from a physical torque, or from an
inertial torque caused by the frame noninertiality, or from nonrigidity
(Getino \& Ferr{\'{a}}ndiz 1990; Escapa, Getino \& Ferr{\'{a}}ndiz 2001,
2002). The free-spin Hamiltonian, expressed through the Euler angles and their
conjugate momenta, is independent of one of the angles, which reveals an
internal symmetry of the problem. In fact, this problem possesses an even
richer symmetry (Deprit \& Elipe 1993), whose existence indicates that the
unperturbed Euler-Poinsot dynamics can be reduced to one degree of freedom.
The possibility of such reduction is not readily apparent and can be seen only
under certain choices of variables. These variables, in analogy with the
orbital mechanics, are called rotational elements. It is convenient to treat
the forced-rotation case as a perturbation expressed through those elements.

The Andoyer variables are often chosen as rotational elements (Andoyer 1923,
Giacaglia \& Jefferys 1971, Kinoshita 1972), though other sets of canonical
elements have appeared in the literature (Richelot 1850; Serret 1866; Peale
1973, 1976; Deprit \& Elipe 1993; Fukushima \& Ishizaki 1994)\footnote{~Some
authors use the term ``Serret-Andoyer elements." This is not correct, because
the set of elements introduced by Richelot (1850) and Serret (1862) differs
from the one employed by Andoyer (1923).}. After a transition to rotational
elements is performed within the undisturbed Euler-Poinsot setting, the next
step is to extend this method to a forced-rotation case. To this end, one will
have to express the torques via the elements. On completion of the
integration, one will have to return back from the elements to the original,
measurable, quantities -- i.e., to the Euler angles and their time
derivatives.

\subsection{The Kinoshita-Souchay theory of rigid-Earth rotation}

The Hamiltonian approach to spin dynamics has found its most important
application in the theory of Earth rotation. A cornerstone work on this topic
was carried out by Kinoshita (1977) who switched from the Euler angles
defining the Earth orientation to the Andoyer variables, and treated their
dynamics by means of the Hori (1966) and Deprit(1969) technique.\footnote{~For
an introduction into the Hori-Deprit method see Boccaletti \& Pucacco (2002)
and Kholshevnikov (1975, 1985). Kinoshita (1977) referred only to the work by
Hori (1966).} Then he translated the results of this development back into the
language of Euler's angles and provided the precessional and nutation
spectrum. Later his approach was extended to a much higher precision by
Kinoshita \& Souchay (1990) and Souchay, Losley, Kinoshita \& Folgueira
(1999).

\subsection{Subtle points}

When one is interested only in the orientation of the rotator, it
is sufficient to have expressions for the Euler angles as
functions of the elements. However, when one needs to know also
the instantaneous angular velocity, one needs expressions for the
Euler angles' time derivatives. This poses the following question:
{\textbf{if we write down the expressions for the Euler angles'
derivatives via the canonical elements in the free-spin case, will
these expressions stay valid under perturbation?}} In the parlance
of orbital mechanics, this question may be formulated like this:
are the canonical elements always osculating? As we shall
demonstrate below, under angular-velocity-dependent disturbances
the condition of osculation is incompatible with that of
canonicity, and therefore expression of the angular velocity via
the canonical elements will, under such types of perturbations,
become nontrivial.

In 2004 the question acquired a special relevance to the Earth-rotation
theory. While the thitherto available observations referred to the orientation
of the Earth figure (Kinoshita et al. 1978), a technique based on ring laser
gyroscope provided a direct measurement of the instantaneous angular velocity
of the Earth relative to an inertial frame (Schreiber et al. 2004, Petrov
2007).

Normally, rotational elements are chosen to have evident physical
interpretation. For example, the Andoyer variable $\,G\,$
coincides with the absolute value of the body's spin angular
momentum, while two other variables, $\,H\,$ and $\,L\,$, are
chosen to coincide, correspondingly,  with the $\,Z$-component of
the angular momentum in the inertial frame, and with its
$\,z$-component in the body frame. The other Andoyer elements,
$\,g\,,\,{\it l}\,,\,h\,$, too, bear some evident meaning. Hence
another important question: {\textbf{will the canonical rotational
elements preserve their simple physical meaning also under
disturbance?}}

 \section{The canonical perturbation theory\\ in orbital and attitude dynamics}

 \subsection{Kepler and Euler}

 In orbital dynamics, a Keplerian conic, emerging as an undisturbed two-body orbit, is
 regarded to be a ``simple motion," so that all the other available motions are
 conveniently considered as distortions of such conics, distortions implemented through
 endowing the orbital constants $C_j$ with their own time dependence. Points of the
 orbit can be contributed by the ``simple curves" either in {\underline{\textbf{a}}}
 nonosculating fashion, as in Fig. 1, or in {\underline{\textbf{the}}} osculating manner,
 as in Fig. 2.

 The disturbances, causing the evolution of the motion from one instantaneous conic to
 another, are the primary's nonsphericity, the gravitational pull of other bodies, the
 atmospheric

 \pagebreak

 \begin{center}
        \epsfxsize=93mm
        \epsfbox{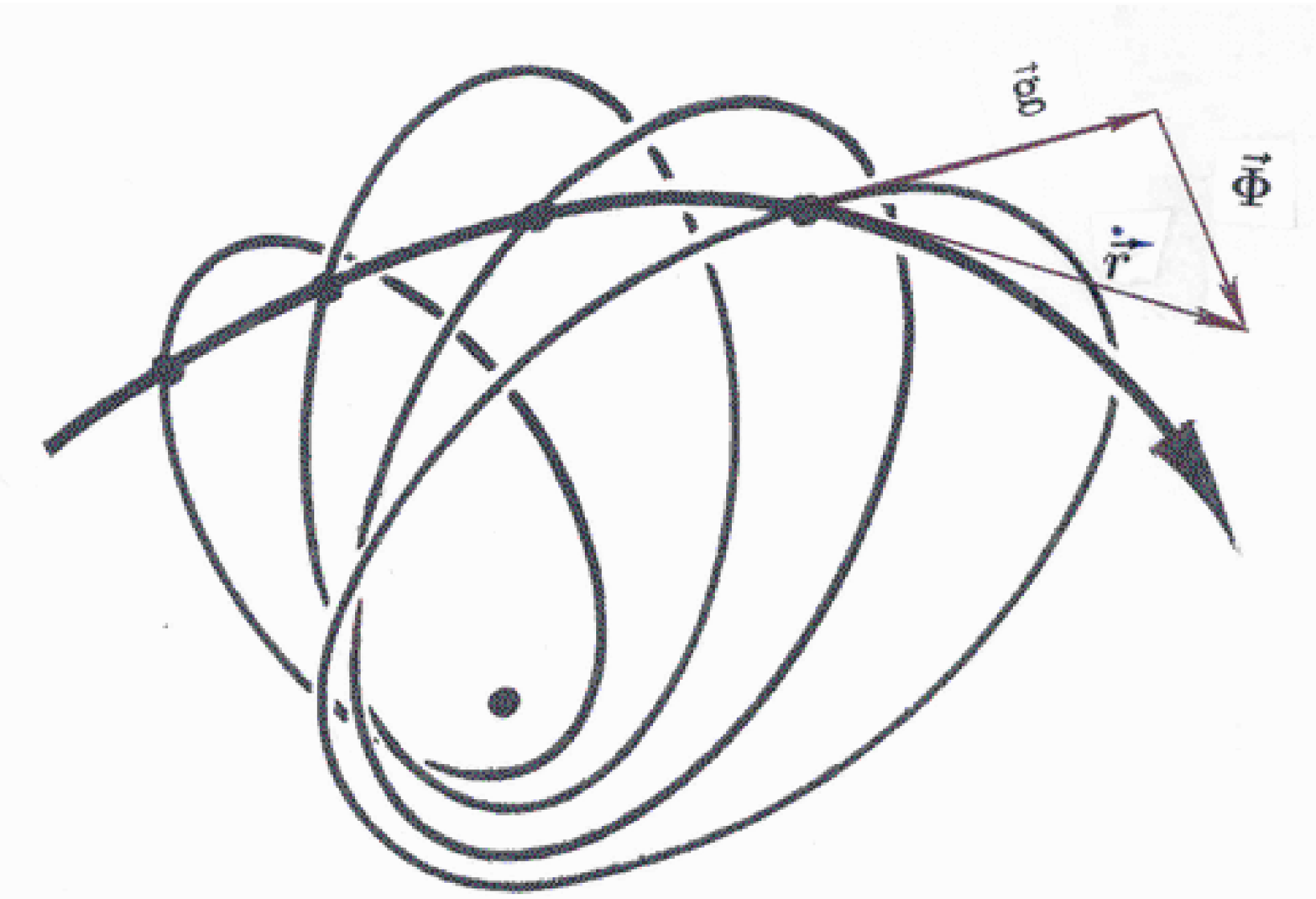}
  \end{center}
  \mbox{\small
 \parbox[b]{5.1in}{{\underline{Fig. 1.}} ~~~ \small The perturbed orbit
 is a set of points belonging to a sequence of confocal instantaneous
 ellipses that are \textbf{not} supposed to be tangent or even coplanar
 to the orbit. As a result, the
physical velocity $\,\doterbold\,$ (tangent to the orbit) differs
from the Keplerian velocity $\,\bf\vec g\,$ (tangent to the
ellipse). To parameterise the depicted sequence of nonosculating
ellipses, and to single it out of the other sequences, it is
suitable to employ the difference between $\doterbold $ and
$\bf\vec g $, expressed as a function of time and six
(nonosculating) orbital elements:
 $\;
 \Phibold(t\,,\,C_1\,,\,.\,.\,.\,,\,C_6)\,=\,
 \doterbold(t\,,\,C_1\,,\,.\,.\,.\,,\,C_6)\,-\,
 {\bf\vec g}(t\,,\,C_1\,,\,.\,.\,.\,,\,C_6)\,.\,
 $
 In the literature,  $\Phibold(t\,,\,C_1\,,\,.\,.\,.\,,\,C_6)$ is
 called gauge function or gauge velocity or, simply, gauge.
 }}

 ~\\

 ~\\

 ~\\

 \begin{center}
        \epsfxsize=93mm
        \epsfbox{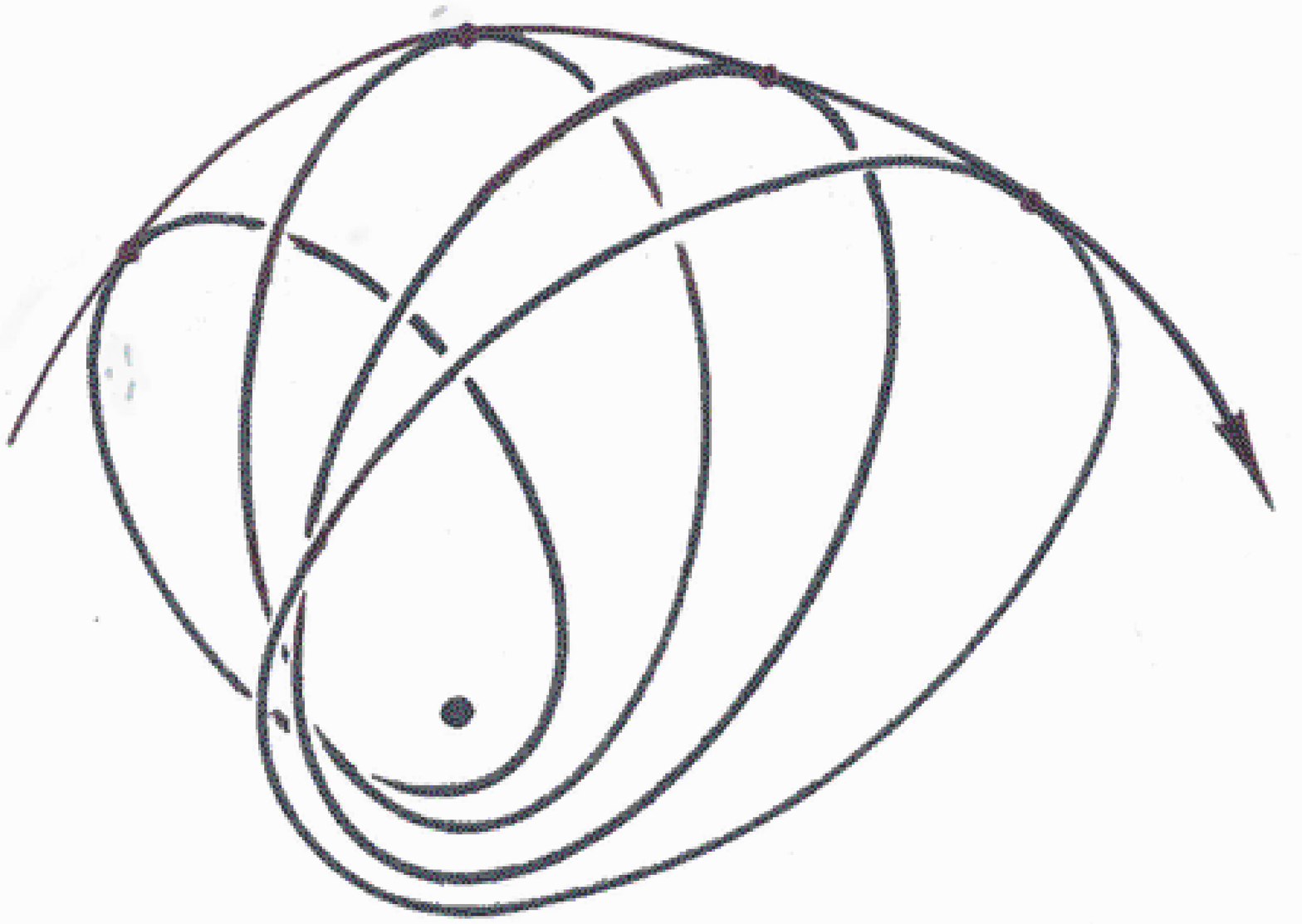}
 \end{center}
  \mbox{\small
 \parbox[b]{5.1in}{{\underline{Fig. 2.}}~~~The orbit
 is represented by a sequence of confocal instantaneous ellipses
 that are
 tangent to the orbit, i.e., osculating. Now, the physical velocity
$\,\doterbold\,$ (tangent to the orbit) coincides with the
Keplerian velocity $\,\bf\vec g\,$ (tangent to the ellipse), so
that their difference vanishes everywhere:
$\,\Phibold(t\,,\,C_1\,,\,.\,.\,.\,,\,C_6)\,=\,0\,$.
 This is the so-called Lagrange constraint or Lagrange gauge. Orbital elements obeying it are
 called
 osculating.
}}

 \noindent
 and radiation-caused drag, the relativistic corrections, and the non-inertiality of
 the reference system.

 On Fig. 1 the orbit consists of points, each of which is donated by a representative of
 a certain family of ``simple" curves (confocal ellipses). These instantaneous ellipses
 are \textbf{not} supposed to be tangent or even coplanar to the orbit. As a result, the
 physical velocity $\,\doterbold \,$ (tangent to the orbit) differs from the Keplerian
 velocity $\,\bf\vec g\;$ (tangent to the ellipse). To parameterise the depicted sequence of nonosculating ellipses, and to
single it out of the other sequences, it is suitable to employ the difference between
$\doterbold$ and $\bf\vec g$, expressed as a function of the time and the orbital
elements:
 $\;
 \Phibold(t\,,\,C_1\,,\,.\,.\,.\,,\,C_6)\,=\,
 \doterbold(t\,,\,C_1\,,\,.\,.\,.\,,\,C_6)\,-\,
 {\bf\vec g}(t\,,\,C_1\,,\,.\,.\,.\,,\,C_6)\;.\;\;
 $
 Evidently,
 \ba
 \nonumber
 \doterbold\,=\,\frac{\partial \erbold}{\partial
 t}\,+\,\sum_{j=1}^{6}\frac{\partial C_j}{\partial
 t}\;\dot{C}_j\;=\;{\bf\vec g}\;+\;\Phibold\;\;,
 \ea
i.e., the unperturbed Keplerian velocity is $\;{\bf\vec
g}\,\equiv\,{\partial \erbold}/{\partial t}\;$, while the said
difference $\Phibold $ is the convective term that emerges when the
instantaneous ellipses are being gradually altered by the
perturbation (and when the orbital elements become time-dependent):
$\,\Phibold\,=\,\sum \left(\partial \erbold/\partial C_j
\right)\,\dot{C}_j\,$. When one fixes a particular functional
dependence of $\,\Phibold\,$ upon time and the elements, this
function, $\,\Phibold(t\,,\,C_1\,,\,.\,.\,.\,,\,C_6)\,$, is called
gauge function or gauge velocity or, simply, gauge.

On Fig. 2, the perturbed orbit is represented with a sequence of
confocal instantaneous ellipses that are tangent to the orbit, i.e.,
osculating. Under this choice, the physical velocity
$\,\doterbold\,$ (tangent to the orbit) will coincide with the
Keplerian velocity $\,\bf\vec g\,$ (tangent to the ellipse), so that
their difference $\;\Phibold(t\,\;C_1\,,\;.\,.\,.\,,\;C_6)\;$ will
vanish everywhere:
 \ba
 \nonumber
 \Phibold(t,\,C_1\,,\,.\,.\,.\,,\,C_6) \equiv\,
 \doterbold(t\,,\,C_1\,,\,.\,.\,.\,,\,C_6) -
 {\bf\vec g}(t\,,\,C_1\,,\,.\,.\,.\,,\,C_6) =\,\sum_{j=1}^{6}\frac{\partial C_j}{\partial
 t}\,\dot{C}_j =0\,.
 \ea
 This, so-called Lagrange constraint or Lagrange gauge, is the necessary and sufficient
 condition of osculation of the orbital elements $\,C_j\,$ (Brouwer \& Clemence 1961).
 Historically, the first attempt of using nonosculating elements dates back to Poincare (1897),
 though he never explored them from the viewpoint of a non-Lagrange constraint choice. (See
 also Abdullah \& Albouy (2001), p. 430.) Parameterisation of nonosculation through a
 non-Lagrange constraint was offered in Efroimsky (2002a,b).

 Similarly to orbital dynamics, in attitude dynamics, a complex spin can be presented as a
 sequence of instantaneous configurations borrowed from a family of some ``simple rotations".
 (Efroimsky 2004) It is convenient to employ in this role the motions exhibited by an
 undeformable free top experiencing no torques.\footnote{~Here one opportunity will be to
 utilise in the role of ``simple" motions the non-circular Eulerian cones described by the
 actual triaxial top, when it is unforced. Another opportunity will be to use, as ``simple"
 motions, the circular Eulerian cones described by a dynamically symmetrical top (and to treat
 its actual triaxiality as another perturbation). The main result of our paper will be
 invariant under this choice.} Each such undisturbed ``simple motion" will be a trajectory on
 the three-dimensional manifold of the Euler angles (Synge \& Griffith 1959). ~For the lack of
 a better term, ~we shall call these unperturbed motions ``Eulerian cones," implying that the
 loci of the rotational axis, which correspond to each such non-perturbed spin state, make
 closed cones (circular, for an axially symmetrical rotator; and elliptic for a triaxial one).
 Then, to implement a perturbed motion, we shall have to go from one Eulerian cone to another,
 just as in Fig. 1 and 2 we go from one Keplerian ellipse to another. Hence, similar to those
 pictures, a smooth ``walk" over the instantaneous Eulerian cones may be osculating or
 nonosculating.

The torques, as well as the actual triaxiality of the top and the
non-inertial nature of the reference frame, will then act as
perturbations causing this ``walk." Perturbations of the latter
two types depend not only upon the rotator's orientation but also
upon its angular velocity.\footnote{
 When we study the Earth rotation relative to the precessing plane of the Earth orbit
 about the Sun, the frame precession gives birth to a fictitious torque (sometimes
 called ``inertial torque") that depends
 upon the Earth's angular velocity.}

\subsection{Delaunay and Andoyer}

In orbital dynamics, we can express the Lagrangian of the reduced
two-body problem via the spherical coordinates
$\;q_j\;=\;\{\,r\,,\;\varphi\,,\;\theta\,\}\;$, then derive their
conjugated momenta $\;p_j\;$ and the Hamiltonian $\;{\cal
H}(q,\,p)\;$, and then carry out the Hamilton-Jacobi procedure
(Plummer 1918), to arrive at the Delaunay variables
 \ba
 \nonumber
\{\,Q_1\,,\;Q_2\,,\;Q_3\;;\;P_1\,,\;P_2\,,\;P_3\,\}\,\equiv\,
\{\,L\,,\;G\,,\;H\;;\;{\it l}_o\,,\;g\,,\;h\,\}\;=~~~~~~~~~\\
\label{1}\\
 \nonumber
\{\,\sqrt{\mu a}\;\;,\;\;\sqrt{\mu a \left(1\,-\,e^2
\right)}\;\;,\;\;\sqrt{\mu a \left(1\,-\,e^2 \right)}\;\cos
\inc\;\;;\;\;-\,M_o\;\;,\;\;-\,\omega\;\;,\;\;-\,\Omega \,\} \;,
\ea
 where $\;\mu\;$ denotes the reduced mass.

Similarly, in rotational dynamics one can define a state of a
spinning top by the three Euler angles
$\;q_j\,=\,\{\,\varphi\,,\;\theta\,,\;\psi\,\}\;$ and their
canonical momenta
$\;p_j\,=\,\{\,p_{\varphi}\,,\,p_{\textstyle{_\theta}}\,,\,p_{\psi}\,\}\;$;
and then carry out a canonical transformation to the Andoyer
elements\footnote{~In attitude dynamics, the Andoyer elements
$\,{\it l}\,,\,g\,,\,h\,$ play the role of coordinates, while
$\,L\,,\,G\,,\,H\,$ are their conjugate momenta. In the orbital
case, the Delaunay variables $\,L\,,\,G\,,\,H\,$ play the role of
coordinates, while $\,{\it l}\,,\,g\,,\,h\,$ defined as in (\ref{1})
act as momenta. Needless to say, this is merely a matter of
convention. (See formulae (9.31 - 9.32) in Goldstein 1981.) For
example in some textbooks on orbital mechanics the Hamiltonian
perturbation is deliberately introduced with an opposite sign, while
the Delaunay elements $\,{\it l}\,,\,g\,,\,h\,$, too, are defined
with signs opposite to given in (\ref{1}). Under such a convention,
the Delaunay elements $\,{\it l}\,,\,g\,,\,h\,$ become coordinates,
while $\,L\,,\,G\,,\,H\,$ act as momenta.} $\{\,{\it
l}\,,\,g\,,\,h\,;\,L\,,\,G\,,\,H\,\}$. By definition, the element
$~G~$ is the magnitude of the
 angular-momentum vector, $~L~$ is the projection of the angular-momentum
 vector on the principal axis $~{\bf{\hat{b}}}_3~$ of the body, while $~H~$
 is the projection of the angular-momentum vector on the $~{\bf{\hat{s}}}_3~
 $ axis of the inertial coordinate system. The variable
 $h~$ conjugate to $~H~$ is the angle from the inertial reference longitude to the ascending
 node of the invariable plane (the one perpendicular to the angular
 momentum). The variable $~g~$ conjugate to $~G~$ is the angle from the
 ascending node of the invariable plane on the reference plane to the
 ascending node of the equator on the invariable plane. Finally, the
 variable conjugate to $~L~$ is the angle $~l~$ from the ascending node of
 the equator on the invariable plane to the the $~{\bf\hat{b}}_1~$ body axis.
 Two auxiliary quantities defined through
 \ba
 \nonumber
 \cos I\;=\;\frac{H}{G}\;\;\;\;,\;\;\;\;\;\;\cos J\;=\;\frac{L}{G}\;\;\;,
 \label{}
 \ea
 have obvious meaning: $~I~$ is the angle between the
 angular-momentum vector and the $~{\bf\hat{s}}_3~$ space axis, while $~J~$
 is the angle between the angular-momentum vector and the $~{\bf\hat{b}}_3~$
 principal axis of the body, as depicted on Fig.~3.

 Andoyer (1923) introduced his variables in a manner different from canonical constants:
 while his variables $\,G\,,\,H\,,\,h\,$ are constants (for a free triaxial rotator),
 the other three, $\,L\,,\,{\it l}\,,\,g\,$, do evolve in time, because the Andoyer
 Hamiltonian of a free top
 \pagebreak
  \begin{center}
        \epsfxsize=177mm
        \epsfbox{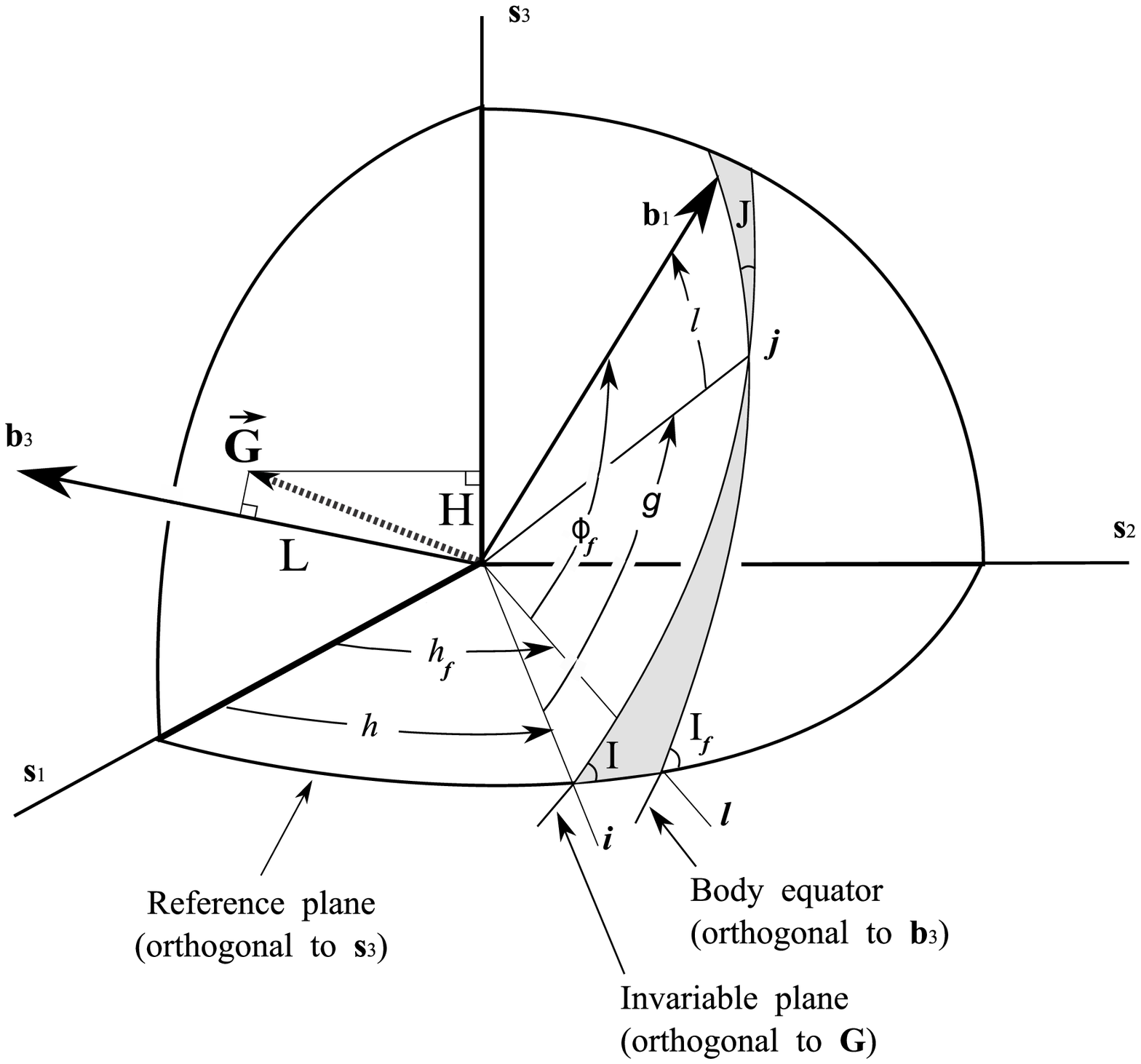}
  \end{center}
  \mbox{\small
  \parbox[b]{5.24in}{{\underline{Fig. 3.}}
  \small{~~A reference coordinate system (inertial or precessing)
  is constituted by axes
  $\,{\bf{s}}_1\,,\,{\bf{s}}_2\,,\,{\bf{s}}_3\,$. A body-fixed
  frame is defined by the principal axes
  $\;{\bf{b}}_1\,,\,{\bf{b}}_2\,,\,{\bf{b}}_3\,$.
  The third frame is constituted by the
  angular-momentum vector $\,{\vec{\textbf{G}}}\,$ and a
  plane orthogonal thereto (the so-called invariable plane).
  The lines of nodes are
  denoted with $\,{\bf{i}},\,{\bf{l}},\,{\bf{j}}\,$. The
  attitude of the body relative to the reference frame is
  given by the Euler angles $\,h_f\,,\;I_f\,,\;\phi_f\,$.
  The orientation of the invariable plane with respect to
  the reference frame is determined by the angles $\,h\,$ and
  $\,I\,$. The inclination $\,I\,$ is equal to the angle that
  the angular-momentum vector $\,{\vec{\textbf{G}}}\,$ makes
  with the reference axis $\,{\bf{s}}_3\,$.
  The angle $\,J\,$ between the invariable plane and the body
  equator coincides with the angle
  that $\,{\vec{\textbf{G}}}\,$ makes with the major-inertia
  axis $\,{\bf{b}}_3\,$ of the body. The projections of the
  angular momentum toward the reference axis $\,{\bf{s}}_3\,$
  and the body axis $\,{\bf{b}}_3\,$ are $\,H\,=\,G\,\cos I\,$
  and $\,L\,=\,G\,\cos J\,$.}}}

 \noindent
  \ba
 \nonumber
 {\mathcal H}(g,\,h,\,l,\,G,\,H,\,L)\,=\,\frac{1}{2}\;
 \left(\frac{\sin^2 l}{A}+\frac{\cos^2 l}{B}\right)\,\left(G^2\,
 -\,L^2\right)\;+\;\frac{L^2}{2\;C}
 \ea
 is a nonvanishing function of the variables $\,{\it l}\,,\,L\,$ and $\,G\,$.
 (Notations $\,A,\,B,\,C\,$ stand for the inertia matrix' principal values that are
 assumed, without loss of generality, to obey the inequality $\,A\,\leq\,B\,\leq\,C\,
 $.) So, to make our analogy complete, we may
 carry out one more
 canonical transformation, from the regular Andoyer set
 $\,\{\,{\it l}\,,\,g\,,\,h\,,\,L\,,\,G\,,\,\,H\,\}\,$ to the ``modified Andoyer set"
 $\,\{\,{\it l}_o\,,\,g_o\,,\,h\,;\,L_o\,,\,G\,,\,H\,\}\,$, where
 $\,L_o\,,\,{\it l}_o\,,\,g_o\,$ are the initial values of $\,L\,,\,{\it l}\,,\,g\,$.
 The modified set consists only of constants of integration, wherefore the appropriate
 Hamiltonian becomes nil.\footnote{~We shall not write down the explicit form of this
transformation, because it is sufficient for us to know that it is canonical. This
follows from the group property of canonical transformations and from the fact that the
transformations from $\;\{\,{\it l}\,,\,g\,,\,h\,,\,L\,,\,G\,,\,H\;\}\;$ to
$\,\{\,\varphi\,,\,\theta\,,\,\psi\,,\;p_{\varphi}\,,\,p_{\theta}\,,\,p_{\psi}\;\}\;$
and from $\;\{\,{\it l}_o\,,\;g_o\,,\;h\,;\;L_o\,,\;G\,,\;H\,\}\;$ to $\;\{\;{\it
l}\,,\,g\,,\,h\,,\,L\,,\,G\,,\,H\;\}\;$ are canonical. The latter transformation is
canonical, for it is simply the time evolution. This canonical transition from
Andoyer-type variables to their initial values is not new -- see Fukushima \& Ishizaki
(1994). Historically, the first set of rotational elements was constituted by constants
(Richelot 1850). Serret (1866) found the generating function of a canonical
transformation from
$\{\,\varphi\,,\,\theta\,,\,\psi\,,\,p_{\varphi}\,,\,p_{\theta}\,,\,p_{\psi}\,\}$ to
that set. His development was polished by Radau (1869) and Tisserand (1889). The
Serret-Richelot set consisted of the following constants:
$\{\,g_o\,,\,h\,,\;-\,t_o\,;\,G\,,\, H\,,\,T_{kin}\,\}$, where $\,h\,,\,G\,$ and $\,H\,$
coincide with the appropriate Andoyer elements, $\,T_{kin}$ is the rotational kinetic
energy, $\,t_o$ is the initial moment of time, and $\,g_o$ is the initial value of the
Andoyer element $\,g$.} Therefore, the modified Andoyer set of variables is analogous to
the Delaunay set with $\,{\it l}_o\,=\,-\,M_o\,$, while the regular Andoyer elements are
analogous to the Delaunay elements with $\,{\it l}\,=\;-\,M\,$ used instead of $\,{\it
l}_o\,=\;-\,M_o\,$. We would stress that, in analogy with the orbital case, the
variables $\,h\,,\,G\,,\,H\,$ are constants (and $\,h_o=h\,,\,G_o=G\,, \,H_o=H$) only in
the unperturbed, free-spin, case.

 To summarise this section, in both cases we start out with
  \ba
 \dot{q}\;=\;\frac{\partial {\cal H}^{(o)}}{\partial p}\;\;\;,\;\;\;\;\;~~
 \dot{p}\;=\;-\;\frac{\partial {\cal H}^{(o)}}{\partial q}
 ~~~,~~~~~~~~~~~~~~~~~~~~~~~~~~~~~~~
 \label{2}
 \ea
$q\;$ and $\;p\;$ being the coordinates and their conjugated
momenta, in the orbital case, or the Euler angles and their
momenta, in the rotation case. Then we switch, via a canonical
transformation
 \ba
 \nonumber
 q\;=\;f(Q\,,\;P\,,\;t)\;\,,\;\\
 \label{3}\\
 \nonumber
 p\;=\;\chi(Q\,,\;P\,,\;t)\;\,,\;\;
 \ea
 to
 \ba
 \dot{Q}\;=\;
 \frac{\partial {\cal H}^*}{\partial P}\;=0\;\;\;,\;\;\;\;\;
 \dot{P}\;=\;-\;\frac{\partial {\cal H}^*}{\partial Q}\;=\;0
 \;\;\;,\;\;\;\;
 {\cal H}^*\;=\;0\;\;,
 \label{4}
 \ea
 where $\;Q\;$ and $\;P\;$ denote the set of Delaunay elements, in the
 orbital case, or the (modified, as explained above) Andoyer
 set $\;\{\,{\it l}_o\,,\;g_o\,,\;h\,;\;L_o\,,\;G\,,\;H\,\}\;$, in
 the case of rigid-body rotation.

This scheme relies on the fact that, for an unperturbed Keplerian
orbit (and, similarly, for an undisturbed Eulerian cone), its
six-constant parameterisation may be chosen so that:\\
~~ \textbf{\underline{1.}}~~~the parameters are constants and, at
the same time, are canonical variables $\,\{\,Q\,,\,P\,\}\,$ with
a zero Hamiltonian:
$\,{\cal H}^*(Q,\,P)\,=\,0\,$; \\
~~\textbf{\underline{2.}}~~$\,$for constant $\,Q\,$ and $\,P\,$,
the transformation equations (\ref{3}) are mathematically
equivalent to the dynamical equations (\ref{2}).

 In practice, this scheme is implemented via the Hamilton-Jacobi
 procedure.

\subsection{Canonical perturbation theory: ~canonicity versus osculation}

Under perturbation, the ``constants" $Q,P$ begin to evolve so that,
after their insertion into
 \ba
 \nonumber
 q\;=\;f\left(\,Q(t)\,,\;P(t)\,,\;t\,\right)\;\;,\\
 \label{5}\\
 \nonumber
 p\;=\;\chi(\,Q(t)\,,\;P(t)\,,\;t\,)\;\;\;\,\;
 \ea
($f$ and $\chi$ being the same functions as in (\ref{3})$\,$), the
resulting motion obeys the disturbed equations
  \ba
 \dot{q}\;=\;\frac{\partial \left({\cal H}^{(o)}\,+\,\Delta {\cal H} \right)}{\partial p}
 \;\;\;, \;\;\;\;\;~~
 \dot{p}\;=\;-\;\frac{\partial \left({\cal H}^{(o)}\,+\,\Delta {\cal H} \right)}{\partial q}
 ~~~.~~~~~~~~~~~~~~~
 \label{6}
 \ea
We also  want our ``constants" $\;Q\;$ and $\;P\;$ to remain
canonical and to obey
  \ba
 \dot{Q}~=~\frac{\partial \left({\cal H}^*\,+\,\Delta {\cal H}^*\right)}{\partial P}\;\;\;,
 \;\;\;\;\;~~
 \dot{P}\;=\;-\;\frac{\partial \left({\cal H}^*\,+\,\Delta {\cal H}^* \right)}{\partial Q}
 ~~~,~~~~~~~~~~~~~~
 \label{7}
 \ea
 where
 \ba
 {\cal H}^*\,=\;0\;\;\;\;\mbox{and}\;\;\;\;\;\Delta {\cal
 H}^*\left(Q\,,\;P\,\;t\right)\;=\;\Delta {\cal
 H}\left(\,q(Q,P,t)\,,\;p(Q,P,t)\,,\;t\,\right)\;\;\;.
 \label{8}
 \ea
Above all, we wish that the perturbed ``constants"
$\,C_j\,\equiv\,Q_1\,,\,Q_2\,,\,Q_3\,,\,P_1\,,\,P_2\,,\,P_3\;$ (the
Delaunay elements, in the orbital case, or the modified Andoyer
elements, in the rotation case) remain osculating. This means that
the perturbed velocity will be expressed by the same function of
$\,C_j(t)\,$ and $\,t\,$ as the unperturbed one used to. Let us
check to what extent this optimism is justified. The perturbed
velocity reads
 \ba
 \dot{q}\;=\;\mbox{g}\;+\;\Phi ~~~,~~~~~~~~~~~~~~~~~~~~~~~~~~
 \label{9}
 \ea
where
 \ba
 \mbox{g}(C(t),\,t)\;\equiv\;\frac{\partial q(C(t),\,t)}{\partial t}
 \;\;~~~~~~~~~~~~
 \label{10}
 \ea
is the functional expression for the unperturbed velocity; and
 \ba
 \Phi(C(t),\,t)\;\equiv\;\sum_{j=1}^6\,\frac{\partial q(C(t),\,t)}{\partial
 C_j}\;\dot{C}_j(t)\;
 \label{11}
 \ea
is the convective term. Since we chose the ``constants" $\,C_j\,$ to
make canonical pairs $\,(Q,\,P)\,$ obeying (\ref{7} - \ref{8}) with
vanishing $\,{\cal H}^*\,$, then insertion of (\ref{7}) into
(\ref{11}) will result in
 \ba
 \Phi\;=\;\sum_{n=1}^3\,\frac{\partial q}{\partial
 Q_n}\;\dot{Q}_n(t)\;+\;\sum_{n=1}^3\,\frac{\partial q}{\partial
 P_n}\;\dot{P}_n(t)\;=\;\frac{\partial \Delta {\cal H}(q,\,p)}{\partial
 p}\;\;\;.
 \label{12}
 \ea
We see that in some situations the canonicity requirement is incompatible with
osculation.\footnote{~For the first time, this observation was made in
Efroimsky \& Goldreich (2003).} To be specific, under a momentum-dependent
perturbation we still can use ansatz (\ref{5}) for calculation of the
coordinates and momenta, but cannot impose the osculation condition
$\,\Phi=0\,$ (i.e., we cannot use $\dot{q}=\mbox{g}$ for calculating the
velocities). Instead, we must use (\ref{9}) with the substitution (\ref{12}).
This generic rule applied both to orbital and rotational motions. Its
application to the orbital case is illustrated by Fig. 2. There, the constants
$\,C_j=(Q_n,\,P_n)\,$ parameterise instantaneous ellipses which, for nonzero
$\,\Phi\,$, are \textbf{not} tangent to the trajectory. In orbital mechanics,
the variables preserving canonicity at the cost of osculation are called
``contact elements" (term coined by Victor Brumberg). The osculating and
contact variables coincide when the disturbance is velocity-independent.
Otherwise, they differ already in the first order of the time-dependent
perturbation (Efroimsky \& Goldreich 2003, 2004). Luckily, in some situations,
their secular parts differ only in the second order (Efroimsky 2005), a
fortunate circumstance anticipated by Goldreich (1965), who came across these
elements in a totally different context unrelated to canonicity.

The case of rotational motion will parallel the theory of orbits.
Now, instead of the instantaneous Keplerian conics, one will deal
with instantaneous Eulerian cones (i.e., with the loci of the
rotational axis, corresponding to non-perturbed spin states).
Indeed, the situation of an axially symmetric unsupported top at
each instant of time is fully defined by the three Euler angles
$\,q_n\,=\,\phi\,,\,\theta\,,\,\psi\,$ and their time derivatives
$\,\dot{q}_n\,=\,\dot{\phi}\,,\,\dot{\theta}\,,\,\dot{\psi}$. The
evolution of these six quantities is governed by three dynamical
equations of the second order (the three projections of $\,d{\bf\vec
L}/dt\,=\,{\taubold}\,$, where $\,{\bf\vec L}\,$ is the angular
momentum and $\,{\taubold}\,$ is the torque) and, therefore, this
evolution will depend upon the time and the six integration
constants:
 \ba
 \nonumber
 q_n\;=\;f_n\left(C_1\,,\;.\,.\,.\,,\;C_6\,\,,\;t
 \right)\;\;\;,\\
 \label{13}\\
 \nonumber
 \dot{q}_n\;=\;\mbox{g}_n\left(C_1\,,\;.\,.\,.\,,\;C_6\,\,,\;t
 \right)\;\;\;,\;
 \ea
where the functions $\,\mbox{g}_n\,$ and $\,f_n\,$ are
interconnected via $\,\mbox{g}_n\,\equiv\,{\partial f_n}/{\partial
t}\,$, for $\,n\,=\,\psi\,,\,\theta\,,\,\phi$.

 Under disturbance, the motion will be altered:
  \ba
 \nonumber
 q_n\;=\;f_n\left(C_1(t)\,,\;.\,.\,.\,,\;C_6(t)\,\,,\;t
 \right)\;\;\;,~~~~~~~~~~~~~~~~~~~~~~~~~~~~~~~~~~~~~~\\
 \label{14}\\
 \nonumber
 \dot{q}_n\;=\;\mbox{g}_n\left(C_1(t)\,,\;.\,.\,.\,,\;C_6(t)\,\,,\;t
 \right)\;+\;\Phi_n\left(C_1(t)\,,\;.\,.\,.\,,\;C_6(t)\,\,,\;t
 \right)\;\;\;,\;
 \ea
where
 \ba
\Phi_n\left(C_1(t)\,,\;.\,.\,.\,,\;C_6(t)\,\,,\;t
 \right)\;\equiv\;\sum_{j=1}^{6}\frac{\partial f_n}{\partial
 C_j}\;\dot{C}_j\;\;\;.
 \label{15}
 \ea
If we want the ``constants" $\,C_j\,$ to constitute canonical pairs
$\,(Q,\,P)\,$ obeying (\ref{7} - \ref{8}), then insertion of
(\ref{7}) into (\ref{15}) will result in
 \ba
 \Phi_n\left(C_1(t)\,,\;.\,.\,.\,,\;C_6(t)\,\,,\;t
 \right)\;\equiv\;\sum \frac{\partial f_n}{\partial
 Q}\;\dot{Q}\;+\;\sum \frac{\partial f_n}{\partial
 P}\;\dot{P}\;=\;\frac{\partial \Delta{\cal H}(q,\,p)}{\partial p_n}\;\;\;,
 \label{16}
 \ea
 so that the canonicity requirement (\ref{7} - \ref{8}) violates
 the gauge freedom in a non-Lagrange fashion.

 To draw this subsection to a close, let us sum up two facts. First,
 no matter what the Hamiltonian perturbation is to be, the Delaunay
 (in the orbital case) or the modified Andoyer (in the attitude case)
 variables $\,Q\,,\,P\,$ always remain canonical. They do so simply
 because they are {\emph{a priori}} defined to be canonical -- see
 equations (\ref{4}) and (\ref{7} - \ref{8}) above. Second, as we
 have seen from (\ref{15} - \ref{16}), the osculating character of
 the $\,Q\,,\,P\,$ variables is lost under momentum-dependent
 perturbations of the Hamiltonian.\footnote{~It is possible, of
 course, to choose the other way and preserve osculation at the
 cost of canonicity. In the orbital case, one should simply set
 $\,\Phibold\,=\,0\,$ in equations
 (52 - 57) of Efroimsky (2006). In the attitude case, though,
 this will be a more cumbersome construction, never implemented
 in the literature hitherto.}

 \subsection{From the modified Andoyer elements to the regular ones}

So far our description of perturbed spin, (\ref{13} - \ref{16}), has
merely been a particular case of the general development (\ref{5} -
\ref{12}). The sole difference was that the role of
canonically-conjugated integration constants $\,C\,=\,(Q,\,P)\,$ in
(\ref{13} - \ref{16}) should be played not by the Delaunay variables
(as in the orbital case) but by some rotational elements -- like,
for example, the Richelot-Serret variables (see the footnote in
subsection 2.2 above) or by the modified Andoyer set $\,(\,{\it
l}_o\,,\,g_o\,,\,h\,;\;L_o\,,\,G\,,\,H\,)\,$ consisting of the
initial values of the regular Andoyer elements. The developments
conventionally used in the theory of Earth rotation, as well as in
spacecraft attitude engineering, are almost always set out in terms
of the regular Andoyer elements, not in terms of their initial
values (the paper by Fukushima \& Ishizaki (1994) being a unique
exception). Fortunately, all our gadgetry, developed above for the
modified Andoyer set, stays applicable for the regular set. To prove
this, let us consider the unperturbed parameterisation of the Euler
angles $\,q_n\,=\,\left(\,\phi\,,\,\theta\,,\,\psi\,\right)\,$ via
the regular Andoyer elements $\,A_j\;=\;(\,{\it
l}\,,\;g\,,\;h\,;\;L\,,\,G\,,\,H\,)\,$:
 \ba
 q_n\;=\;f_n\left(\;A_1(C\,,\,t)\;,\;.\,.\,.\;,\;A_6(C\,,\,t)\;\right)\;\;\;,
 ~~~~~~~~~~~~~~~~~~~~~~
 \label{17}
 \ea
each element $\,A_i$ being a function of time and of the initial
values $\,C_j = ({\it l}_o\,,\,g_o\,,\,h\,;\,L_o\,,\,G\,,\,H\,)\,$.
When a perturbation gets turned on, the parameterisation (\ref{17})
stays, while the time evolution of the elements $\,A_i\,$ changes:
beside the standard time-dependence inherent in the free-spin
Andoyer elements, the perturbed elements acquire an extra
time-dependence through the evolution of their initial
values.\footnote{~This is fully analogous to the transition from the
unperturbed mean longitude,
 \ba
 \nonumber
 M(t)\;=\;M_o\;+\;n\; \left( \,t\;-\;t_o\, \right)\;\;\;,\;\;\;\;\;{\mbox{with}}\;\;\;\;
 M_o\,,\;n\,,\;t_o\;=\;const\;\;\;,
 \ea
 to the perturbed one,
 \ba
 \nonumber
 M(t)~=~M_o(t)~+~\int_{t_o}^{t}~n(t')~dt'\;\;\;,\;\;\;\;{\mbox{with}}\;\;\;\;
 t_o\,=\;const\;\;\;,\;\;\;\;\;\;\;
 \ea
in orbital dynamics.
 }
 Then the time evolution of an Euler angle
$\;q_n\,=\,\left(\,\varphi\,,\;\theta\,,\;\psi\,\right)\,$ will be
given by a sum of two items: (1) the angle's unperturbed
dependence upon time and time-dependent elements; and (2) an
appropriate addition $\;\Phi_n\;$ that arises from a
perturbation-caused alteration of the elements' dependence upon
the time:
 \ba
 \dot{q}_n\;=\;\mbox{g}_n\;+\;\Phi_n~~~.~~~~~
 \label{18}
 \ea
 The unperturbed part is
 \ba
 \mbox{g}_n\,=\,\sum_{i=1}^{6}\;\frac{\partial f_n}{\partial A_i}\;
 \left(\frac{\partial A_i}{\partial t}\right)_C\;\;\;,
 \label{19}
 \ea
 while the convective term is given by
 \ba
 \nonumber
 \Phi_n =\sum_{i=1}^{6}\sum_{j=1}^{6}\left(\frac{\partial f_n}{\partial A_i}\right)_t
 \left(\frac{\partial A_i}{\partial C_j}\right)_t\,\dot{C}_j =
 \sum_{j=1}^{6}\left(\frac{\partial f_n}{\partial
 C_j}\right)_t\,\dot{C}_j~~~~~~~~~~~~~~~
 \ea
 \ba
 =~
 \sum_{j=1}^{3}\left(\frac{\partial f_n}{\partial Q_j}\right)_t\,\dot{Q}_j
 ~+~
 \sum_{j=1}^{3}\left(\frac{\partial f_n}{\partial
 P_j}\right)_t\,\dot{P}_j\,=\;\frac{\partial \Delta{\cal H}(q,p)}{\partial p_n}\;\;,\;
 \label{20}
 \ea
where the set $\;C_j\;$ is split into canonical coordinates and momenta like
this: $\;Q_j\;=\;(\,{\it l}_o\,,\,g_o\,,\,h\,)\;$ and
$\;P_j\;=\;(\,L_o\,,\,G\,,\,H\,)\,$. In the case of free spin they obey the
Hamilton equations with a vanishing Hamiltonian and, therefore, are all
constants. In the case of disturbed spin, their evolution is governed by
(\ref{7} - \ref{8}), substitution whereof in (\ref{20}) will once again take
us to (\ref{16}). This means that the non-osculation-caused convective
corrections to the velocities stay the same, no matter whether we parameterise
the Euler angles through the modified Andoyer elements (variable constants) or
through the regular Andoyer elements. This invariance will become obvious if,
once again, we consider the analogy with orbital mechanics: on Fig. 1, the
correction $\;\Phibold\;$ is independent of how we choose to parameterise the
nonosculating instantaneous ellipse -- through the Delaunay set
with $\;M_o\;$ or through the one containing $\;M\;$.\\

This consideration yields the following consequences:\\

{{\textbf{(a)}}} ~~Under momentum-dependent perturbations, calculation of the
angular velocities via the elements must be performed not through the second
equation of (\ref{13}) but through the second equation of (\ref{14}), with
(\ref{16}) substituted therein. The convective term given by (\ref{16}) is nonzero
when the perturbation is angular-velocity-dependent. In other words, under such
type of perturbations, the canonicity condition imposed upon the Richelot-Serret or
the Andoyer elements is incompatible with osculation. An example of such
perturbation shows itself in the theory of planetary rotation, when we switch to a
coordinate system associated with the orbit plane. Precession of this plane makes
the frame noninertial, and the appropriate Lagrangian perturbation depends upon the
planet's angular velocity. The corresponding Hamiltonian perturbation (denoted in
the Kinoshita-Souchay theory by $\,E\,$) comes out momentum-dependent. In this
theory the Andoyer elements are introduced in the precessing frame, and since the
precession-caused perturbation is momentum-dependent, these elements come out
nonosculating. For this reason, their substitution into the undisturbed expressions
(2.6) and (6.26 - 6.27) in Kinoshita (1977) will not render the angular velocity
relative to the precessing frame wherein the elements were introduced. To furnish
the angular velocity relative to that frame, these expressions must be amended with
the appropriate convective terms.
 \\

{{\textbf{(b)}}} ~~The above circumstance, instead of being a flaw of the
Kinoshita-Souchay theory, turns out to be its strong point. It can be shown that
Kinoshita's undisturbed expressions for the angular velocity via the elements keep
rendering the {\emph{inertial}} angular velocity, even when the elements defined in
a precessing frame are plugged therein. Briefly speaking, we first introduce the
Andoyer elements in an unperturbed setting (inertial frame) and write down the
expressions, via these elements, for the Euler angles and velocities relative to
the inertial frame. Then we introduce a momentum-dependent perturbation, i.e.,
switch to a precessing frame, and in that frame we introduce the Andoyer elements.
Insertion thereof into the unperturbed expressions for the Euler angles and angular
velocities gives us the Euler angles relative to the precessing frame and (due to
the nonosculating nature of the elements) the angular velocity relative to the
{\emph{inertial}} frame, not to the precessing one.\footnote{~This mishap is an
example of osculation loss. We introduce the elements in a certain frame (the
precessing frame of the orbit), plug them into the unperturbed expressions for the
Euler angles and for the angular velocities, and here comes the result: while we
obtain the correct values of the Euler angles relative to the said frame, we do
{\bf{not}} get the right values for the angular velocity relative to that frame.
(Instead, our formulae return the values of the angular velocity relative to
another, inertial, frame.) This happens because the disturbance, associated with a
transition to the precessing frame, depends not only upon the Earth's orientation
but also upon its angular velocity. Or, stated alternatively, because the
appropriate Hamiltonian variation $\,\Delta {\cal H}\,$ depends upon the momenta
$\,p\,$ canonically conjugated to the Euler angles $\,q\,$:
 \ba
 \nonumber
 \Phi\,\equiv\,\frac{\partial {\Delta{\cal H}}}{\partial
 p}\,\neq\,0\;\;\;.
 \ea
 } A proof of this fact will be presented in Appendix 1.3.

This fact should not be regarded as a disadvantage of the Kinoshita-Souchay theory,
because in some situations it is the inertial, not the relative, angular velocity
that is measured (Schreiber et al. 2004, Petrov 2007). Under these circumstances,
Kinoshita's expressions for the angular velocity should be employed (as long as
they are correctly identified as the formulae for the {\emph{inertial}} angular
velocity).


 \section{The angular velocity relative to the precessing frame}

 In the theory of Earth rotation, three angular velocities emerge:
  \ba
 \nonumber
 \omegabold^{^{(rel)}}\;\equiv\;\; \;\mbox{the relative angular
 velocity,}\,~~~~~~~~~~~~~~~~~~~~~~~~~~~~~~~~~~~~~~~~~~~~~~~~~~\,~~~~\\
 \nonumber
 \mbox{~~~~~~~i.e., the body's angular velocity relative to a
 precessing orbital frame;~}\\
 \nonumber\\
 \nonumber
 \mubold\;\,\;\;\;\;\equiv\;\,\;\mbox{the precession rate of the orbital frame with
 respect to an inertial one;}\\
 \nonumber\\
 \nonumber
 \omegabold^{^{(inert)}}\equiv\;\;\,\mbox{the inertial angular
 velocity,}~~~~~~~~~~~~~~~~~~~~~~~~~~~~~~~~~~~~~~~~~~~~~~~~~~~~~~~\\
 \nonumber
 \mbox{~~~~~~~i.e., the body's angular velocity with respect to the inertial frame.~~~~~}
 \ea
 Evidently, the latter is the sum of the two former ones:
 $\;\omegabold^{^{(inert)}}=\,\omegabold^{^{(rel)}}+\,\mubold\;$.

 If some day we develop an experimental technique for measuring the Earth's angular
 velocity relative to the precessing plane of its orbit, we shall have to compare
 the observations with the theoretical predictions for the directional angles of this,
 {\emph{relative}}, angular velocity $\,\omegabold^{^{(rel)}}\,$.

 The Kinoshita (1977) theory was created with intention to furnish the
 precessing-frame-related directional angles\footnote{~Here and hereafter the term
 ``directional angles" will stand for the longitude of the node and the inclination
 of the plane perpendicular to the Earth figure. An analogous meaning is understood
 for the directional angles of the angular velocity.}
 of the Earth figure (formulae (2.3) and (6.24 - 6.25) in Kinoshita's paper). This
 theory also provides precessing-frame-related directional angles of the
 Earth's angular-velocity vector (formulae (2.6) and (6.26 - 6.27) in
 {\emph{Ibid.}}). We prove in Appendix 1.3 below that, contrary to the
 expectations, the latter expressions render the directional angles not
 of the relative but of the {\emph{inertial}} angular velocity
 $\;\omegabold^{^{(inert)}}\;$:
 \ba
 I_r^{^{(inert)}}~=~I~+~J~\left(1\,-\,\frac{C}{2A}\,-\,\frac{C}{2B}\right)\,\left[\,
 \cos {g}\,-\,{\it e}\,\cos \left(2\,{\it
 l}+{g}\right)\,
 \right]~~~~,~~~~~~~~~~~~~~~~~~~~~~~~~~~
 \label{548}
 \label{421}
 \ea
 and
 \ba
 h_r^{^{(inert)}}\,=\;h\;+\;\frac{J}{\sin \,I}\;\left(1\,-\,\frac{C}{2A}\,
 -\,\frac{C}{2B}\right)\;\left[\;\sin {g}\;-\;{\it e}\;\sin \left(2\;{\it
 l}\;+\;{g}\right)\;\right]\,~~~.~~~~~~~~~~~~~~~~~~~~~
 \label{551}
 \label{422}
 \ea
 where the angles $\,I\,$ and $\,J\,$ are as on Fig. 3, while $\;e\;$ is introduced as a
 measure of triaxiality of the rotator:\footnote{~For the Earth, $\,J\,\sim\,10^{-6}$
 rad, which justifies the common approximation to write all formulae up to the first
 order in $\,J\,$ (Kinoshita 1977). The value of the triaxiality parameter is: $\;e\,=\,
 3.3646441\,\times\,10^{-3}\;$ (Escapa, Getino \& Ferr{\'{a}}ndiz 2002).}
 \ba
 e\;\equiv\;\frac{\left[\;\left(1/B\right)\;-\;\left(1/A\right)\;\right]/2}{
 \left(1/C\right) \;-\;\left[\left(1/A\right)\;+\;\left(1/B\right)\right]/2}\;\;\;
 \label{50}
 \label{423}
 \ea
 $A\,,\;B\,,\;C$ being the principal moments of
 inertia.\footnote{~The Earth is assumed to be rigid, and its body axes
 are chosen to diagonalise its inertia matrix.}

 This is a very nontrivial and counterintuitive fact. On
 introducing the Andoyer variables in the precessing frame of
 orbit, we plug them into the standard expressions for the
 orientation angles and the angular velocity. Doing so, we
 naturally expect to obtain the orientation and the spin rate
 relative to that precessing frame. We indeed get the body
 orientation relative to that frame, but the rendered angular
 velocity turns out to be {\bf{not}} the one relative to the
 precessing frame wherein the Andoyer elements were introduced.
 Instead, the standard formulae give us the angular velocity
 relative to some other frame, the inertial one (as if we had
 used the Andoyer variables defined in the inertial frame).
 This is an interesting (and still underappreciated by
 mathematicians) internal symmetry instilled into the Andoyer
 construction: we can go through a continuum of Andoyer sets
 (each set introduced in a different precessing frame), but
 their substitution into the standard formulae for the angular
 velocity will always return the angular velocity relative to
 the inertial frame.

 A proof of this fact begins with a study of the physical meaning of the Andoyer elements
 introduced in a precessing frame (presented in Appendix A.1.1). Completion of the proof
 demands a sequence of calculations so laborious that we chose to put them into the
 Appendix (see Appendices A.1.2 - A.1.3). This entire situation remarkably parallels a
 similar episode from the theory of Delaunay elements in orbital dynamics (see the end of
 Appendix A.1.3).

 Now, what if we want to know the angular velocity relative to
 the precessing frame, i.e., $\,\omegabold^{^{(rel)}}\;$?
 The precessing-frame-related directional angles of this angular
 velocity will look as
 \ba
 I_r^{^{(rel)}}~=~I_r^{^{(inert)}}~+\;{I_r}^{^{(\Phi)}}~~~,~~~~~~~~~~~~~~~~~~~~~~~~~~~
 \label{48}
 \label{424}
 \ea
 and
 \ba
 h_r^{^{(rel)}}\,=\;h_r^{^{(inert)}}~+\,h_r^{^{(\Phi)}}~~~,~~~~~~~~~~~~~~~~~~~~~~~~~~~
 \label{51}
 \label{425}
 \ea
 ~\\
 the extra terms $\,{I_r}^{^{(\Phi)}}\,,\;{h_r}^{^{(\Phi)}}\,$ emerging
 because, as explained above, one has to add the convective
 term $\;\Phi\;$ to the unperturbed velocity $\;\mbox{g}\;$, in order to obtain
 the full velocity $\;\dot{q}\;$ under disturbance. Here $\;q\;$ stands
 for the three Eulerian angles\footnote{~Be mindful that in the physics
 and engineering literature the Euler angles are traditionally denoted
 with $\,(\,\phi\,,\,\theta\,,\,\psi\,)\,$. In the literature on the
 Earth rotation, very often the inverse convention,
 $\,(\,\psi\,,\,\theta\,,\,\phi\,)\,$, is employed. In the
 Kinoshita-Souchay theory, these angles are denoted with
 $\,(\,h\,,\,I\,,\,\phi\,)\,$. The angles defining orientation of the
 Earth's figure are accompanied with the subscript $\,f\;$ and are
 termed as: $\,(\,h_f\,,\,I_f\,,\,\phi_f\,)\,$. The directional angles
 of the Earth's angular-velocity vector are equipped with subscript
 $\,r\,$. For the relative and the inertial angular velocities the
 angles are denoted with $\,(\,h_r^{^{(rel)}}\,,\,I_r^{^{(rel)}}\,,\,
 \phi_r^{^{(rel)}}\,)\,$ and $\,(\,h_r^{^{(inert)}}\,,\,
 I_r^{^{(inert)}}\,,\,\phi_r^{^{(inert)}}\,)\,$, accordingly.}
 $\,q_n\,=\,\{\,h_f\,,\,I_f\,,\,\phi_f\,\}\,$
defining the the orientation of the principal axes of the Earth,
relative to the precessing frame, so $\,\dot{q}_n\,$ will signify
time derivatives of Euler angles relative to this precessing frame.
The convective terms, entering the expressions for
$\,\dot{q}_n\,=\,\{\,\dot{h}_f\,,\,\dot{I}_f\,,\,\dot{\phi}_f\,\}\,$,
can be calculated using formula (\ref{16}) -- see the Appendices 3
and 4 below. The ensuing corrections to the Euler angles determining
the orientation of the instantaneous spin axis will look as
 \ba
 \nonumber
 I_r^{^{(\Phi)}}\,=\;
 -\,\dot{\pi}_1\;\frac{C}{L}\;\cos I\;\sin (h-\Pi_1) \,+\,\dot{\Pi}_1\;
 \frac{C}{L}\,
 \left[\,\sin \pi_1\;\cos I\;\cos (h-\Pi_1)\,+\;\cos\pi_1\;\sin I\,-\;\sin
 I\,\right]
 ~~~~~~~\\
 \nonumber\\
 +\,O\left(J^2\right)\,+\,O(J\Phi/\omega)\,+\,O(\,(\Phi/\omega)^2\,)\;\;\;\;\;
 \label{49}
 \label{426}
 \label{926}
 \ea
 and
 \ba
 h_r^{^{(\Phi)}}\,=\;-\;\dot{\pi}_1\;\frac{C}{L}\;\,\frac{\cos (h-\Pi_1)}{\sin I}
 \,-\,\dot{\Pi}_1\;\,\frac{C}{L}\;\;\frac{\sin\pi_1\;\sin (h-\Pi_1)}{\sin
 I}\,+\,O(J^2)\,+\,
 O(\,(\Phi/\omega)^2\,)\,+\,O(\,J\,\Phi/\omega\,)\;\;.~~
 \label{52}
 \label{427}
 \label{927}
 \\
 \nonumber
 \ea
 These corrections depend upon two angles that define the orientation of the precessing
 orbit with respect to an inertial frame -- the inclination $\;\pi_1\;$ and the node
 $\;\Pi_1\,$.

 Let us make rough numerical estimates for the case of the rigid Earth.
 Putting together (\ref{421}), (\ref{424}), and (\ref{426}),
 we see that in the resulting expression for $\,I_r^{^{(rel)}}\,$
 \ba
 \nonumber
 I_r^{^{(rel)}}=\,I\,+\,J\,\left(1\,-\,\frac{C}{2A}\,-\,\frac{C}{2B}\right)\,\left[
 \cos {g}\,-\,{\it e}\,\cos \left(2\,{\it
 l}+{g}\right) \right]\,-\,
 %
 %
 \dot{\pi}_1\;\,\frac{C}{L}\,\;\cos I\;\sin (h-\Pi_1)  ~~~~~~~~~~~~~\\
 \nonumber\\
 +\,\dot{\Pi}_1\;\frac{C}{L}\,\left[\,\sin \pi_1\;\cos I
 \;\cos (h-\Pi_1)\,+\,\cos\pi_1\;\sin I\,-\,\sin I\,\right]
 \,+\,O(J^2)\,+\,O(\,J\,\Phi/\omega\,)\,+\,O(\,(\Phi/\omega)^2\,)~~~~~
 \label{I_r}
 \ea
 we have the leading term,
 $\,I\,$, and three additions -- ~of order $\;J\,\sim\,10^{-6}\;$, of
 order $\;J\,e\,\sim\,10^{-9}\;$, and of order\footnote{~This estimate ensues from the
 trivial observation that $\,C/L\,\approx\,\omega^{-1}\,$. Regarding the numbers:
 according to Lieske et al. (1977) and Seidelmann (1992), $\dot{\pi}_1 \sim
 47"/century$, while
 $\dot{\Pi}_1 \sim -870"/century\approx$ $-\,2.4\times10^{-3}deg/yr$. On the
 other hand, $\omega\sim360\,deg/day\sim1.3\times10^{5}deg/yr\,$ whence
 $\;\Phi/\omega \sim \dot{\pi}_1/\omega \sim \,10^{-9}$.
 (We could as well
 have used the IERS value of $\,\omega\,\approx\,7.3\,\times\,10^{-5}\,rad/s\,\sim\,
 1.3\,\times\,10^{7}\,deg/century\,\sim\,4.7\,\times\,10^{10}\;"/century\,$.)}
 $\,\Phi/\omega\,\sim\,\dot{\pi}_1/\omega\,\sim\,10^{-9}$.
 We see that {\emph{the nonosculation-caused convective terms
 $\,\Phi\,$ provide an effect on the spin-axis orientation, which is
 of the same order as the $\,J\,e\,$ term stemming from
 triaxiality}}.\footnote{~The nutational spectra of these two
 contributions are, however, quite different (secular vs. periodic).}
 As  $\;1\,rad\,\approx\,0.2\,\times\,10^{\,6}\;''\,$ and
 $\;J\,\sim\,10^{-6}\;$, then the $\,J\,$ term brings into $\,I_r\,$
 a contribution of an arcsecond order, while the $\,\Phi\,$ and $\,Je\,$
 terms give corrections of order milliacrseconds. We also see that the terms of
 order $\;J\;\Phi/\omega\,$ and those of $\;(\Phi/\omega)^2\;$ are much less than
 one percent of a microarcsecond and may be neglected.

 Numerical estimates for the expression
 \ba
 \nonumber
 h_r^{^{(rel)}}\;=\;h\;+\;\frac{J}{\sin \,I}\;\left(1\,-\,\frac{C}{2A}\,
 -\,\frac{C}{2B}\right)\;\left[\;\sin {g}\;-\;{\it e}\;\sin \left(2\;{\it
 l}\;+\;{g}\right)\;\right] ~~~~~~~~~~~~~~~~~~~~~~~~~~~~~~
 \ea
 \ba
 ~~~~~~-\;\dot{\pi}_1\;\frac{C}{L}\;\frac{\cos (h-\Pi_1)}{\sin I}\;-\;
 \dot{\Pi}_1\;\frac{C}{L}\;\,\frac{\sin\pi_1\;\,\sin (h-\Pi_1)}{\sin
 I}\,+\,O(J^2)\,+\,
 O(\,(\Phi/\omega)^2\,)\,+\,O(\,J\,\Phi/\omega\,)\;\;~~
 ~~~
 \ea
 will be similar.

 Formulae (\ref{426} - \ref{427}) constitute the main result of this
 paper.\footnote{~Through the medium of equations (\ref{460} -
 \ref{462}) it is also possible to express these corrections via the
 Euler set, instead of the Andoyer variables.} In Appendices 2 - 4 we
 present their derivation based on formulae (\ref{18}) and
 (\ref{20}). It would be important to note that the resulting
 corrections acquire the form (\ref{51} - \ref{52}) provided the
 coordinate system co-precessing with the orbit is chosen as in
 Kinoshita (1977), i.e., by three consecutive Euler rotations
 $\;{\bf{\hat{R}}}_{3}(\,-\,\Pi_1)\;{\bf{\hat{R}}}_{N}(\pi_1)\;
 {\bf{\hat{R}}}_{Z}(\Pi_1)\;$, letter $\,Z\,$ standing for an
 inertial axis orthogonal to the ecliptic of epoch, $\,N\,$
 denoting the line of nodes, and $\,3\,$ being a precessing axis
 perpendicular to the ecliptic of date -- see Appendix A.2.2.3.
 Under an alternative choice of axes within the co-precessing
 frame, expressions for $\;I_r^{^{(rel)}}\;$ and $\;h_r^{^{(rel)}}
 \;$ will look differently. For example, a transition carried out
 by only two Euler rotations, $\;{\bf{\hat{R}}}_{N}(\pi_1)\;
 {\bf{\hat{R}}}_{Z}(\Pi_1)\;$, as in Appendix A.2.2.2, will yield
 expression (\ref{996}) instead of (\ref{51}), and (\ref{496})
 instead of (\ref{52}).

 In principle, (\ref{51} - \ref{52})
 might as well be derived by purely geometrical means, i.e., from the formula
 $\;\omegabold^{^{(inert)}}=\,\omegabold^{^{(rel)}}+\,\mubold\;$.
 We however chose the method based on (\ref{18}) and (\ref{20}),
 because this method is fundamental and applicable to {\emph{any
 kind}} of momentum-dependent perturbations of the Hamiltonian --
 for example, to the perturbations caused by deviations from
 rigidity, as studied by Getino \& Ferr{\'{a}}ndiz (1994).
 A similar situation will emerge in the (yet to be built)
 relativistic theory of the Earth rotation.
 ~\\

\section{Conclusions}

 In this article we explained that the unperturbed spin states (``Eulerian cones")
 play in the attitude dynamics the same role as the unperturbed two-body orbits
 (``Keplerian conics") play in the orbital mechanics. Just as the orbital elements
 parameterising Keplerian conics, the rotational elements parameterising Eulerian
 cones may be either osculating or nonosculating. If the perturbation depends upon
 the velocity (in the orbital case) or upon the angular velocity (in the
 attitude case), the condition of osculation is incompatible with the condition of
 canonicity. In these situations the standard equations furnish the Delaunay (in
 the orbital case) or Andoyer (in the attitude case) elements, which are not
 osculating, -- circumstance important when the elements are employed for
 calculation of the velocity or angular velocity. The functional form of the
 expression for a velocity or an angular velocity through elements depends
 upon whether these elements are osculating or not.

 A remarkable peculiarity is shared by the Delaunay and Andoyer
 elements. Suppose the perturbation is caused by a transition to a
 precessing frame of reference, and the elements are introduced in
 this noninertial frame. Their substitution into the unperturbed
 expressions for the Cartesian coordinates (or the Euler angles)
 will render the right position (or the attitude) relative to the
 precessing frame wherein these elements were defined. Now, suppose
 that we impose on our elements the condition of canonicity. Since
 the frame-precession-caused perturbation is momentum-dependent,
 the canonicity condition is incompatible with the osculation one.
 Hence, when our elements are inserted into the unperturbed
 expressions for the velocity or angular velocity, they will NOT
 return the velocity with respect to the precessing frame. It turns
 out, though, that they will render the velocity relative to the
 inertial frame. While for the orbital case this was proven in
 Efroimsky \& Goldreich (2003, 2004), in the current paper we
 proved this fact also for the attitude case.

 This has ramifications for the Kinoshita-Souchay theory of the
 Earth rotation. In this theory, the Andoyer elements are defined in
 a precessing frame of the Earth orbit. In Kinoshita (1977) these
 elements were {\emph{ab initio}} canonical -- simply because
 Kinoshita obtained them via a canonical transformation (see section
 3 of his work). As demonstrated in sections 2.3 - 2.4 of our paper,
 the by-default-imposed canonicity condition made the elements
 nonosculating. Insertion of such elements into the
 unperturbed equations for the angular velocity (formulae (2.6) and
 (6.26 - 6.27) in Kinoshita 1977) does not yield the angular
 velocity relative to the frame wherein the elements were defined
 (the precessing frame). Rather, the equations will still furnish
 the angular velocity relative to the inertial frame of reference.
 This way of osculation loss might be a flaw of the Kinoshita-Souchay theory, had
 we expected it to render the angular velocity with respect to a
 precessing frame. In reality, the osculation loss is an advantage
 of the theory, because the presently available experimental
 technique (Schreiber et al. 2004, Petrov 2007) provides for the measurement of
 the angular velocity relative to the inertial frame -- the
 velocity furnished by the Kinoshita-Souchay theory.

 In the final section we provide expressions (\ref{49}
 - \ref{52}) for the body-frame-related directional angles of the planet's
 angular velocity relative to a frame coprecessing with the planet's orbit.
 The method wherewith we calculate these angles is general and
 applicable to to {\emph{any kind}} of momentum-dependent perturbations of
 the Hamiltonian --
 for example, to the perturbations caused by deviations from
 rigidity.


 ~\\

{\underline{\bf{\Large{Acknowledgments}}}}\\
~\\
 The authors would like to deeply thank Hiroshi Kinoshita for his extremely valuable consultations on some subtleties of his theory. ME is also grateful
 to Pini Gurfil, George Kaplan, Jean Souchay, and Jack Wisdom for fruitful and stimulating conversations on the subject. AE's contribution was partially supported by Spanish Projects AYA2004-07970 and AYA2005-08109.

 ~\\

 \noindent
{\underline{\bf{\Large{\textbf{Appendix 1.}}}}} \\
~\\
 {\underline{\bf{\Large{\textbf{The Andoyer variables introduced
in a precessing frame}}}}}

  ~\\

 \noindent
 {\large{\textbf{A$\,$1.1 ~~Formalism}}}\\

 Let us consider an unsupported rigid body whose spin is to be studied in a
 coordinate system, which itself is precessing relative to some inertial
 frame. The said system is assumed to precess at a rate $\,\mubold\,$ so
 the kinetic energy of rotation, in the inertial frame, is given by
 \ba
 \nonumber
 T_{kin}\,=\;
 \frac{1}{2}\;{{\omegabold}^{^{(inert)}}}^{^T}{\mathbb I}\;
 {\omegabold}^{^{(inert)}}=\;\frac{1}{2}\;\left({\omegabold}^{^{(rel)}}+\,
 \mubold\right)^{^T}{\mathbb I}\;\left({\omegabold}^{^{(rel)}}+\,\mubold
 \right)\;=\;\frac{1}{2}\;\sum_{i=x,y,z}\;I_i\,\left(\,\omega_i^{^{(rel)}}
 \,+\;\mu_i\,\right)^2\\
 \nonumber\\
 \label{112}
 \label{428}\\
 \nonumber
 =\,
 \frac{1}{2}\;A\,\left(\,\omega_x^{^{(rel)}}\,+\;\mu_x\,\right)^2\,+\,
 \frac{1}{2}\;B\,\left(\,\omega_y^{^{(rel)}}\,+\;\mu_y\,\right)^2\,+\,
 \frac{1}{2}\;C\,\left(\,\omega_z^{^{(rel)}}\,+\;\mu_z\,\right)^2\;\;\;,\;
 \;~~~~~~~~~~~~~
 \ea
 where $\,I_i\,\equiv\,(A\,,\,B\,,\,C)\,$ are the principal values of the
 inertia matrix of the body, $~\omegabold^{^{(inert)}}$ is the inertial
 angular velocity (i.e., the one with respect to an inertial frame),
 $~\omegabold^{^{(rel)}}$ is the relative angular velocity (i.e., the one
 with respect to a precessing coordinate system), while $~\mubold~$ is the
 rotation rate of the precessing frame with respect to the inertial frame.
 In (\ref{112}), both $~\omegabold\,$'{\small{s}~} and $~\mubold~$ are
 resolved into their components along the principal axes $\;{\hat{\bf b}}_1
 \equiv\,{\hat{\bf{x}}}\,,\;{\hat{\bf b}}_2\equiv\,{\hat{\bf{y}}}\,,\;
 {\hat{\bf b}}_3\equiv\,{\hat{\bf{z}}}\;$ of the rotating body. Expression
 (\ref{112}) is fundamental and stays, no matter whether $\,\mubold\,$
 depends on the rotator's orientation, or whether it carries a direct time
 dependence.

 The role of canonical coordinates will be played the Euler
 angles\footnote{~We would once again remind that the Euler angles, though
 normally termed $\,(\,\phi\,,\,\theta\,,\,\psi\,)\,$, in the astronomical
 literature are often denoted as $\,(\,\psi\,,\,\theta\,,\,\phi\,)\,$. In
 the Kinoshita-Souchay theory notations $\,(\,h\,,\,I\,,\,\phi\,)\,$ are
 employed. The angles defining orientation of the Earth's figure and of the
 Earth's angular-velocity vector are accompanied with the subscripts $f$
 and $r$, correspondingly: $\,(\,h_f\,,\,I_f\,,\,\phi_f\,)\,$ and $\,(\,
 h_r\,,\,I_r\,,\,\phi_r\,)\,$. The directional angles of the inertial
 angular velocity will be denoted with  $\,(\,h_r^{^{(inert)}}\,,\,
 I_r^{^{(inert)}}\,,\,\phi_r^{^{(inert)}}\,)\,$. Those of the relative
 velocity will be called $\,(\,h_r^{^{(rel)}}\,,\,I_r^{^{(rel)}}\,,\,
 \phi_r^{^{(rel)}}\,)\,$.}
 \ba
 q_n~=~(\,h_f\,,\;I_f\,,\;\phi_f\,)
 \label{429}
 \ea
 that map the {\emph{precessing}} coordinate basis into the principal body
 basis. To compute their conjugate momenta, let us assume that
 noninertiality of the precessing coordinate system is the only
 angular-velocity-dependent perturbation. Then the momenta are simply the
 derivatives of the kinetic energy. With aid of the formulae for the
 body-frame components of the relative angular velocity,\footnote{~It
 should be emphasised, that the components of the angular velocity
 $\,\omegabold^{^{(rel)}}$ are related to the body axes, but the angular
 velocity itself is the relative one (i.e., that with respect to the
 precessing coordinate system). Our formulae (\ref{430} - \ref{432})
 are analogous to equations (2.4) in Kinoshita (1977). At the initial step
 of his development, Kinoshita used his equations (2.4) to express the
 inertial angular velocity (what we call $\,\omegabold^{^{(inert)}}\,$) via
 the Euler angles introduced in an inertial frame. Then, on having
 introduced a precessing orbital frame, Kinoshita employed these equations for
 expressing the angular velocity through the Euler angles introduced in a
 precessing frame. Kinoshita did not explore whether this operation would
 furnish the relative angular velocity (what we call
 $\,\omegabold^{^{(rel)}}\,$) or still the inertial one.}
  \ba
 \omega_x^{^{(rel)}}\;=\;\dot{h_{f}}\;\sin I_{f}\;\sin\phi_{f}
 \;+\;\dot{I_{f}}\;\cos\phi_{f} ~~~,~~~~~~~~~
  \label{430}\\
 \nonumber\\
 \omega_y^{^{(rel)}}\;=\;\dot{h_{f}}\;\sin I_{f}\;\cos\phi_{f}
 \;-\;\dot{I_{f}}\;\sin\phi_{f} ~~~,~~~~~~~~\,
 \label{431}\\
 \nonumber\\
 \omega_z^{^{(rel)}}\;=\;\dot{h_{f}}\;\cos I_{f}\;+\;\dot{\phi_{f}}
 \;\;\;,~~~~~~~~~~~~~~~~~~~~~~~~~\,~
  \label{432}
 \ea
 we obtain:
 \ba
 p_{_{h_f}}=\frac{\partial T_{kin}}{\partial\dot{h}_f}=
     A\left(\omega_x^{^{(rel)}}+\mu_x\right)\sin I_f \,\sin \phi_f +
     B\left(\omega_y^{^{(rel)}}+\mu_y\right)\sin I_f \,\cos \phi_f +
     C\left(\omega_z^{^{(rel)}}+\mu_z\right)\cos I_f\;\;,\;\;\;\;\;
 \label{113}
 \label{433}
 \ea
 \ba
 p_{_{I_f}}\;=\;\frac{\partial T_{kin}}{\partial\dot{I}_f\;}\;=\;
     A\;\left(\,\omega_x^{^{(rel)}}\,+\;\mu_x\right)\;\cos \phi_f\; -\;
     B\;\left(\,\omega_y^{^{(rel)}}\,+\;\mu_y\right)\;\sin
     \phi_f\;\;\;.~~~~~~~~~~~~~~~~~~~~~~~~~~~~~~~~\,
 \label{114}
 \label{434}
 \ea
  \ba
 p_{_{{\phi}_f}}\;=\;\frac{\partial
 T_{kin}}{\partial\dot{\phi}_f\;}\;=\;C\;\left(\,\omega_z^{^{(rel)}}\,+\;
 \mu_z\,\right)\;\;\;\;,\;\;\;\;\;\;\;\;\;\;\;\;\;\;\;\;\;\;\;\;\;\;\;\;\;
 \;\;\;\;\;\;\;\;\;\;\;\;\;\;\;\;\;\;\;\;\;\;\;\;\;\;\;\;\;\;\;\;\;\;\;\;~
 ~~~~~~~~~~~~~~~~~~~
 \label{115}
 \label{435}
 \ea
 These formulae enable one to express the angular-velocity components
 $~\omega_i~$ and the derivatives $~\dot{q}_n\,=\,(\,\dot{h}_f\,,\,
 \dot{I}_f\,,\,\dot{\phi}_f\,)~$ via the momenta $~{p}_n\,=\,(\,p_{_{h_f}}
 \,,\,p_{_{I_f}}\,,\,p_{_{\phi_f}}\,)\,$. Insertion of (\ref{113} -
 \ref{115}) into
 \ba
 {\cal H}\,=\,\sum^{3}_{n=1
 }\;\dot{q}_n\,p_n \,-\,{\cal
 L}\,=\,\dot{h}_f\,p_{_{h_f}}\,+\,\dot{I}_f\,p_{_{I_f}}
 \,+\,\dot{\phi}_f\,p_{_{{\phi}_f}}\,-\,T\,+\,V(h_f\,,\;I_f\,,\;\phi_f\;;\,
 \;t)\;\;\;\;\;\;\;\;\;\;\;\;
 \label{116}
 \label{436}
 \ea
 results, after some tedious algebra, in
 \ba
 {\cal H}\,=\,T\,+\,\Delta{\cal H}\;\;\;,
 \label{117}
 \label{437}
 \ea
 the perturbation $\;\Delta{\cal H}\;$ consisting of a
 potential term $\;V\;$ (presumed to depend only upon the
 time and the angular coordinates, not upon the momenta)
 and a precession-generated inertial term $\,E\,$:
 \ba
 \nonumber
 \Delta{\cal H}\;=\;V(h_f\,,\;I_f\,,\;\phi_f\;;\,\;t)\;+\;E\;\;\;,
 \label{}
 \ea
 where
 \ba
 \nonumber
 E\,=\,-\,\mu_x\,\left[\,\frac{\sin\phi_f}{\sin I_f}\,
 \left(p_{_{h_f}}\,-\,p_{_{\phi_f}}\,\cos
 I_f\right)\,+\,p_{_{I_f}}\,\cos\phi_f\,\right]
 ~~~~~~~~~~~~~~~~~~~~~\\
 \label{118}
 \label{E}
 \label{438}\\
 \nonumber\\
 \nonumber
 -\,\mu_y\,\left[\,\frac{\cos\phi_f}{\sin I_f}\,\left(p_{_{h_f}}\,-\,
 p_{_{\phi_f}}\,\cos I_f\right)\,-\,p_{_{I_f}} \,\sin \phi_f\,\right]
 \, -\,\mu_z\;p_{_{\phi_f}}~~~~,~~~~
 \ea
 expression equivalent to formulae (24 - 25) in Giacaglia \& Jefferys
 (1971).\footnote{~It can be shown (Gurfil, Elipe, Tangren \&
 Efroimsky 2007) that the body-frame-related
 components $\,g_i\,$ of the angular momentum are connected with the Euler angles
 and their conjugate momenta through
 \ba
 \nonumber
 g_1 = p_{_{h_f}}\;\frac{\sin\phi}{\sin I}\,+\,p_{_{I_f}}\,\cos\phi\,-\,
 p_{_{\phi_f}}\,\sin\phi\,\cot I\;\;,~~~~~
 g_2 =p_{_{h_f}}\frac{\cos\phi}{\sin I}-p_{_{I_f}}\,\sin\phi - p_{_{\phi_f}}\,\cos\phi
 \,\cot I~~,~~~~~ g_3 = p_{_{\phi_f}} ~~,
 \ea
 whence it can be seen that (\ref{118}) is merely another form of the relation $\,
 \Delta{\cal{H}}\,=\,-\,\mubold\cdot{\vec{\mbox{\bf{G}}}}\,$. The latter can also be
 expressed via the Andoyer variables:
 \ba
 \nonumber
 \Delta {\cal{H}}\;=\;-\;\mubold\cdot{\vec{\mbox{\bf{G}}}}\;=\;-\;\mu_1\;\sqrt{G^2\,-\;L^2}
 \;\sin{\it{l}}\;-\;\mu_2\;\sqrt{G^2\,-\;L^2}\;\cos{\it{l}}\;-\;\mu_3\;L\;\;\;.~~~~
 \ea}
 Now let us employ the machinery set out in subsection 2.2. The fact that
 the Andoyer elements are introduced in a noninertial frame is accounted
 for by the emergence of the $\,\mu$-terms in the expression (\ref{116})
 for the disturbance $\,\Delta \cal H\,$. Insertion of (\ref{118}) into
 (\ref{20}) entails:
 \ba
 \dot{q}_n\,=\,\mbox{g}_n\,+\,\frac{\partial \Delta \cal H}{\partial p_n}
 \label{119}
 \label{439}
 \ea
 where $\,q_n\,\equiv\,h_f\,,\,I_f\,,\,\phi_f\,$, and the convective terms
 are given by
 \ba
 \frac{\partial \Delta\cal H}{\partial p_{_{h_f}}}\;=\;\,-\;\,\frac{~\mu_x
 \;\sin\phi_f\;+\;\mu_y\;\cos\phi_f~}{\sin I_f}\;\;\;,~~~~~~~~~~~~~
 \label{120}
 \label{440}
 \ea
 \ba
 \frac{\partial \Delta \cal H}{\partial p_{_{I_f}}}\;=\;\,-\;\,\mu_x\;
 \cos\phi_f\;+\;\mu_y\;\sin\phi_f\;\;\;\;\;\,,~~~~~~~~~~~~~
 \label{121}
 \label{441}
 \ea
 \ba
 \frac{\partial \Delta \cal H}{\partial p_{_{\phi_f}}}\;=\;\left(\,\mu_x
 \;\sin\phi_f\;+\;\mu_y\;\cos\phi_f\,\right)\;\cot I_f\;-\;\mu_z~~~,\,
 \label{122}
 \label{442}
 \ea
 where $\,\dot{q}_n\,$ stand for $\,\,\dot{h}_f\,,\,\dot{I}_f\,,\,
 \dot{\phi}_f\,$, and $\,{p}_n\,$ signify the corresponding momenta,
 while $\;\mu_x\,,\;\mu_y\,,\,\mu_z\,$ are the components of $\,\mubold\,$
 in the principal axes of the body.\\

 \noindent
 {\large{\textbf{A$\,$1.2 ~~The physical interpretation of the Andoyer variables\\
 $\left.~~~~~~~~\,\right.$ defined in a precessing frame}}}\\

 The physical content of the Andoyer construction built in an inertial
 frame is transparent: see Fig.~3 and explanation thereto. Will all
 the Andoyer variables and the auxiliary angles $~I~$ and $~J~$ retain the
 same physical meaning if we re-introduce the Andoyer construction in a
 noninertial frame? The answer is affirmative, because a transition to a
 noninertial frame is no different from any other perturbation: precession
 of the fiducial frame $~\left({\bf\hat{s}}_1\,,\,{{\bf\hat{s}}_2}\,,\,
 {{\bf\hat{s}}_3}\right)~$ is equivalent to emergence of an extra
 perturbing torque, one generated by the inertial forces (i.e., by the
 fictitious forces emerging in the noninertial frame of references). In
 the original Andoyer construction assembled in an inertial space, the
 invariable plane was orthogonal to the instantaneous direction of the
 angular-momentum vector: if the perturbing torques were to
 instantaneously vanish, the angular-momentum vector (and the invariable
 plane orthogonal thereto) would freeze in their positions relative to the
 fiducial axes $~\left({\bf\hat{s}}_1\,,\,{{\bf\hat{s}}_2}\,,\,
 {{\bf\hat{s}}_3}\right)~$ (which were inertial and therefore indifferent
 to vanishing of the perturbation). Now, that the Andoyer construction is
 built in a precessing frame, the fiducial plane is no longer inertial.
 Nevertheless if the inertial torques were to instantaneously vanish, then
 the invariable plane would still freeze relative to the fiducial plane
 (because the fiducial plane would seize its precession). Therefore, all
 the variables retain their initial meaning. In particular, the variables
 $\;I\;$ and $\;J\;$ defined as above will be the angles that the
 angular-momentum makes, correspondingly, with the precessing
 $~{\bf\hat{s}}_3~$ space axis and with the $~{\bf\hat{b}}_3~$ principal
 axis of the body.\footnote{~On the interrelation between the Andoyer
 variables, referred to an inertial frame, and those referred to a moving
 frame see equation (3.3) in Kinoshita (1977).} Among other things, this
 explains why Laskar \& Robutel (1993) and Touma \& Wisdom (1993, 1994),
 who explored the history of the Martian obliquity, arrived to very close
 results. Both groups rightly used the angle $\,I\,$ as an approximation
 for the obliquity. While Touma \& Wisdom (1993, 1994) employed (a
 somewhat simplified version of) the Kinoshita formalism in an inertial
 frame, Laskar \& Robutel (1993) used this machinery in a precessing frame
 of the orbit. Now we understand why they obtained so close results, with
 minor differences stemming, most likely, from averaging-caused error
 accumulation in the latter paper. (The computation by Touma and Wisdom
 was based on unaveraged equations of motion, while Laskar and Robutel
 employed orbit-averaged equations.)\\
 ~\\

 \noindent
 {\large{\textbf{A$\,$1.3 ~~Calculation of the angular velocities via the Andoyer\\
  $\left.~~~~~~~~\,\right.$ variables introduced in a precessing frame of reference}}}\\

 Let us now have a look at the well-known expressions
   \ba
   \omega_x^{^{(rel)}} \;= \;\dot{h}_f\;\sin I_f\;\sin \phi_f\;+
   \;\dot{I}_f\;\cos\phi_f\;\;\;,
  \label{omega1}
 \label{443}
  \ea
  \ba
  \omega_y^{^{(rel)}} \;= \;\dot{h}_f \;\cos \phi_f\;\sin I_f\;-
  \;\dot{I}_f\;\sin \phi_f\;\;\;,
  \label{omega2}
 \label{444}
  \ea
  \ba
  \omega_z^{^{(rel)}}\;=\;\dot{\phi}_f\;+\;\dot{h}_f\;\cos
  I_f\;\;\;,~~~~~~~~~~~~~~~~~
  \label{omega3}
  \label{445}
   \ea
 for the principal-axes components of the precessing-frame-related angular
 velocity $\,\omegabold^{^{(rel)}}$. These
 formulae render this angular velocity as a function of the rates of Euler
 angle's evolution, so one can symbolically denote the functional
 dependence (\ref{omega1} - \ref{omega3}) as $\;\omegabold\,=\,\omegabold
 (\dot{q})\;$. This dependence is linear, so insertion of (\ref{119})
 therein will yield:
 \ba
 \omegabold\left(\,\dot{q}(A)\,\right)\;=\;\omegabold(\,\mbox{g}(A)\,)~+~
 \omegabold\left(~{\partial \Delta \cal H}/{\partial p} \;\right)\;\;\;,
 \label{123}
 \label{446}
 \ea
 with $\,A\;$ denoting the set of Andoyer variables, and $\,p\,$ signifying the
 canonical momenta corresponding to the Euler angles. Direct substitution of (\ref{120} - \ref{122})
 into  (\ref{omega1} - \ref{omega3}) will then show that the second term on the right-hand side in
 (\ref{123}) is exactly $\;\;-\,\mubold\;$:
  \ba
 \omegabold(\dot{q}(A)) \;=\;\omegabold(\mbox{g(A)})\;-\;\mubold\;\;\;.
 \label{124}
 \label{447}
 \ea
 Since the $\;\omegabold(\dot{q})\;$ is, {\emph{ab initio}}, the relative angular
 velocity $\,\omegabold^{^{(rel)}}\,$ (i.e., that of the body frame relative to the
 precessing frame), and since $~\mubold~$ is the precession rate of that frame with
 respect to the inertial one, then $\;\omegabold(\,\mbox{g}(A)\,)\;$ will always
 return the inertial angular velocity of the body, $\,\omegabold^{^{(inert)}}\,$
 (i.e., the angular velocity relative to the inertial frame). It will do so even
 despite the fact that now the Andoyer parameterisation is introduced in a precessing
 coordinate frame!

 In brief, the above line of reasoning may be summarised as:
 \begin{eqnarray}
 \left.
 \begin{array}{ccc}
 \omegabold\left(\,\dot{q}(A)\,\right)\;=\;\omegabold(\,\mbox{g}(A)\,)~+~
 \omegabold\left(~{\partial \Delta \cal H}/{\partial p}
 \;\right)\;\;\;,\;\\
 ~\\
 \omegabold(\,\dot{q}(A)\,)\,=\;\omegabold^{^{(rel)}}\;\;,~~~~~~~~~~~~~~~~~
 ~~~~~~~~~~~~~~~\\
 ~\\
\omegabold(\,\partial \Delta{\cal H}/\partial p\,)\,=\,\mubold
~~~,\,~~~~~~~~~~~~~~~~~~~~~~~~~~~~~~\\
~\\
 \omegabold^{^{(rel)}}\,=\;\omegabold^{^{(inert)}}-\,\mubold\,
 ~~~.~~~~~~~~~~~~~~~~~~~~~~~~~~~
 \end{array}
 \right\}\;\;\;\Longrightarrow\;\;\;~~
 \omegabold(\,\mbox{g}(A)\,)\;=\;\omegabold^{^{(inert)}}\;\;\;\;~~~~~~
 \label{448}
 \end{eqnarray}
 where the entities are defined as follows:
 \ba
 \nonumber
 \omegabold^{^{(rel)}}\;\equiv\;\; \;\mbox{the relative angular
 velocity,}\,~~~~~~~~~~~~~~~~~~~~~~~~~~~~~~~~~~~~~~~~~~~~~~~~~~~~~\\
 \nonumber
 \mbox{~~~~~~~i.e., the body's angular velocity relative to a
 precessing orbital frame;}\\
 \nonumber\\
 \nonumber
 \mubold\;\,\;\;\;\;\equiv\;\,\;\mbox{the precession rate of that frame with respect to
 an inertial one;~~~~~~~}\\
 \nonumber\\
 \nonumber
 \omegabold^{^{(inert)}}\equiv\;\;\,\mbox{the inertial angular
 velocity,}~~~~~~~~~~~~~~~~~~~~~~~~~~~~~~~~~~~~~~~~~~~~~~~~~~~~~~\\
 \nonumber
 \mbox{~~~~~~~i.e., the body's angular velocity with respect to the inertial frame.~~~~}
 \ea

 This development parallels a situation in orbital dynamics.
 There the role of canonical elements is played by the Delaunay set
 $\;C\,=\,\left(\,Q\,;\;P\,\right)\,=\,\left(\,L\,,\,G\,,\,H\,;\;-\,M_o\,,
 \,-\, \omega\,,\,-\,\Omega\,\right)\;$ with
 \ba
 \nonumber
 L\,\equiv\,\mu^{1/2}\,a^{1/2}\,\;\;,\;\;\;\;\;
G\,\equiv\,\mu^{1/2}\,a^{1/2}\,\left(1\,-\,e^2\right)^{1/2}\,\;\;,\;\;\;\;
H\,\equiv\,\mu^{1/2}\,a^{1/2}\,\left(1\,-\,e^2\right)^{1/2}\,\cos
\inc\,\;\;\;\;\;,\;\;
 \label{}
 \ea
 the parameters $\,a,\,e,\,i,\,\omega,\,\Omega,\,M_o\,$ being the Kepler
 orbital elements. In the unperturbed setting (the two-body problem in
 inertial axes), the Cartesian coordinates $\,\erbold\,\equiv\,(\,x_1\,,\,
 x_2\,,\,x_3)\,$ and velocities $\,(\,\dot{x}_1\,,\, \dot{x}_2\,,\,
 \dot{x}_3)\,$ are expressed via the time and the Delaunay constants
 by means of the following functional dependencies:
 \ba
 \erbold~=~\efbold\left(\,C\,,~t\,\right)~~~~~\mbox{and}~~~~~{\vec{\bf{\mbox{\bf{v}}}}}~=~
 {\vec{\bf{\mbox{\bf{g}}}}}\left(\,C\,,~t\,
 \right)~~~,\;\;\;\;\mbox{where}\;\;\;{\vec{\bf{\mbox{\bf{g}}}}}\;\equiv\;
 {\partial\efbold}/{\partial t}\;\;\;.
 \label{449}
  \ea
 If we want to describe a satellite orbiting a precessing oblate planet,
 we may fix our reference frame on the precessing equator of date. Then
 the two-body problem will get amended with two disturbances. One, $\;
 \Delta{\cal H}_{oblate}\;$, caused by the presence of the equatorial
 bulge of the planet, will depend only upon the satellite's position.
 Another one, $\;\Delta{\cal H}_{precess}~$, will stem from the
 noninertial nature of our frame and, thus, will give birth to
 velocity-dependent inertial forces. Under these perturbations, the
 Delaunay constants (now introduced in the precessing frame) will
 become canonical variables evolving in time. As
 explained in subsection 2.3, the velocity-dependence of one of the
 perturbations involved will make the Delaunay variables nonosculating
 (provided that we keep them canonical). On
 the one hand, the expression $\;\erbold\;=\;\efbold\left(\,C(t)\,,\;t\,
 \right)\;$ will return the correct Cartesian coordinates of the satellite
 in the precessing equatorial frame, i.e., in the frame wherein the
 Delaunay variables were introduced. On the other hand, the expression
 $~{\vec{\mbox{\bf{g}}}} \left(\,C\,,\;t\,\right)~$ will no longer return the correct
 velocities in that frame. Indeed, according to (\ref{9} - \ref{10}), the
 Cartesian components of the velocity in the precessing equatorial frame
 will be given by $\;{\vec{\mbox{\bf{g}}}} \left(\,C\,,\;t\,\right)\;+\;{\partial \Delta
 {\cal H}_{precess}}/{\partial \vec{\mbox{\bf{p}}} }~$. However, since the second term of this
 sum is equal to $\;-\,\mubold\times\erbold\,$, then
 $\;\vec{\mbox{\bf{g}}} \left(\,C\,,~t\,\right)~$ turns out to always render the
 velocity with respect to the inertial frame of reference (Efroimsky \&
 Goldreich 2004, Efroimsky 2005).\\

 ~\\


 \noindent
{\underline{\bf{\Large{\textbf{Appendix 2.}}}}}\\
~\\
{\underline{\bf{\Large{\textbf{The
instantaneous angular velocity}}}}}\\
~\\
{\underline{\bf{\Large{\textbf{in the Kinoshita-Souchay theory}}}}}\\
~\\

The main burden of subsection 2.3 in the text above was to highlight
the need to add the convective term $\;\Phi\;$ to the unperturbed
velocity $\;\mbox{g}\;$, in order to obtain the full velocity
$\;\dot{q}\;$ under disturbance. Here $\;q\;$ stands for a vector
consisting of the three Eulerian angles
$\,q_n\,=\,h_f\,,\,I_f\,,\,\phi_f\,$ defining the orientation of the
principal axes of the Earth relative to the precessing frame. The
corresponding convective terms, entering the expressions for
$\,\dot{q}_n\,=\,\dot{h}_f\,,\,\dot{I}_f\,,\,\dot{\phi}_f\,$, are
given by formula (\ref{20}). Our eventual goal will be to calculate
the corresponding corrections to the Euler angles determining the
instantaneous axis of rotation in a precessing frame of reference.\\

 \noindent
 {\large{\textbf{A$\,$2.1 ~~The unperturbed velocities}}}\\

 In this subsection we shall write the unperturbed Euler angles'
 partial time derivatives $\;\mbox{g}_n\,\equiv\,{\partial q_n}/
 {\partial t}\;$ as functions of these angles and of the Andoyer
 variables.

 In the Kinoshita-Souchay theory of Earth rotation, the Euler
 angles defining the figure of the Earth are denoted
 with
 \ba
 q_n\;=\;(\,h_f\;,\;I_f\;,\;\phi_f\,)\;\;\;,
 \label{21}
 \label{450}
 \ea
 the subscript standing for ``figure."

 Now, let us denote the principal body axes with $\;1\,,\;2\,,\;3\;$
 and the appropriate moments of inertia with $\;A\,,\;B\,,\;C\;$ (so
 that $\;A\;\leq\;B\;\leq\;C\;$). The angular momentum
 $\,\bf{\vec{\,L}}\,$ is connected with the Earth-figure Euler angles
 via the body-frame components (\ref{omega1} - \ref{omega3}) of the
 inertial-frame-related\footnote{~Be mindful that in formulae
 (\ref{omega1} - \ref{omega3}) the notations $\,h_f\,,\,I_f\,,\,\phi_f
 \,$ stood for the Euler angles defining the body's orientation
 relative to a precessing frame. For this reason, (\ref{omega1} -
 \ref{omega3}) furnished the relative angular velocity
 $\,\omegabold^{^{(rel)}}\,$. In this subsection we are beginning
 with the unperturbed situation, when the orbit frame is yet assumed
 to be inertial. Hence, at this moment, $\,h_f\,,\,I_f\,,\,\phi_f
 \,$ yet denote the angles relative to the inertial frame, and hence
 the same formulae render the inertial angular velocity
 $\,\omegabold^{^{(inert)}}\,$.} angular velocity
 $\,\omegabold^{^{(inert)}}\,$:
 \ba
 L_{x}\;=\;A\;\omega_x^{^{(inert)}}\;=\;A\;\left(\dot{h_{f}}\;\sin I_{f}\;\sin\phi_{f}
 \;+\;\dot{I_{f}}\;\cos\phi_{f} \right)~~~,~~~~~~~~~
  \label{451}\\
 \nonumber\\
 L_{y}\;=\;B\;\omega_y^{^{(inert)}}\;=\;B\;\left(\dot{h_{f}}\;\sin I_{f}\;\cos\phi_{f}
 \;-\;\dot{I_{f}}\;\sin\phi_{f}  \right)~~~,~~~~~~~~\,
 \label{22}
  \label{452}\\
 \nonumber\\
 L_{z}\;=\;C\;\omega_z^{^{(inert)}}\;=\;C\;\left( \dot{h_{f}}\;\cos I_{f}\;+\;\dot{\phi_{f}}
 \right)\;\;\;.~~~~~~~~~~~~~~~~~~~~~~~~~\,~
  \label{453}
 \ea
 On the other hand, the body-frame components of the angular momentum
 will be related to the Andoyer elements through\footnote{~At this
 point, we are discussing the unperturbed case, with no frame
 precession. However, as explained in subsection A1.2, the
 interconnection (\ref{454} - \ref{456}) between the Andoyer elements
 and the components of $\,{\bf{\vec{\,L}}}\,$ will stay valid also
 when the precession is ``turned on" (and both the elements and the
 components of $\,{\bf{\vec{\,L}}}\,$ are introduced in a precessing
 frame of reference).}
 \ba
 L_x\;=\;\sqrt{G^2\;-\;L^2}\;\sin {\it l}\;\;\;,
  \label{454}\\
 \nonumber\\
 L_y\;=\;\sqrt{G^2\;-\;L^2}\;\cos {\it l}\;\;\;,
 \label{23}
  \label{455}\\
 \nonumber\\
 L_z\;\equiv\;L\;\;\;.~~~~~~~~~~~~~~~~~~~\,
  \label{456}
 \ea
Substituting (\ref{454} - \ref{456}) into (\ref{451} - \ref{453}) and solving
for the rates of change of the Euler angles will entail:
 \ba
 \frac{\partial h_f}{\partial t}\,=\,
 \frac{1}{\sin I_{f}}\,\left[\frac{L_x}{A}\,\sin\phi_{f}\,+\,\frac{L_y}{B}\,\cos\phi_{f}
 \right]\,=\,
 \frac{1}{\sin I_{f}}\,\sqrt{G^2\,-\,L^2}\,\left[ \frac{\sin {\it l}\,\sin\phi_{f}}{A}\,+
 \,\frac{\cos {\it l}\,\cos\phi_{f}}{B} \right] \;\;\;,\;\;\;\;
 \label{24}
  \label{457}
 \\
 \nonumber\\
 \nonumber\\
 \frac{\partial I_f}{\partial t}\;=\;\frac{L_x}{A}\;\cos\phi_{f}\;-\;\frac{L_y}{B}\;
 \sin\phi_{f} \; = \;\sqrt{G^2\,-\,L^2}\;
  \left[ \frac{\sin {\it l}\;\cos\phi_{f}}{A}\;-\;\frac{\cos {\it l}\;\sin\phi_{f}}{B}
 \right]~~~,~~~~~~~~~~~~~~~~~~~
 \label{25}
  \label{458}
 \ea
 ~\\
 \ba
 \nonumber
 \frac{\partial {\phi}_f}{\partial t}\,=\,\frac{L_z}{C}\,-\,\cot I_{f}\,
 \left[\frac{L_x}{A}\,\sin\phi_{f}\,+\,\frac{L_y}{B}\,\cos\phi_{f} \right] ~~~~~~~~~~~~
 ~~~~~~~~~~~~~~~~~~~~~~~~~~~~~~~~~~~~~~~~~~~~~~~~~~~\\
 \nonumber\\
 \nonumber\\
 =\,\frac{L}{C}\,-\,\sqrt{G^2\,-\,L^2}\,\cot I_{f}\,
 \left[ \frac{\sin {\it l}\,\sin\phi_{f}}{A}\,+\,\frac{\cos {\it l}\,\cos\phi_{f}}{B}
 \right]\;\;\;,\;\;\;
 \label{26}
  \label{459}
 \ea
where we deliberately replaced
$\;\dot{h_{f}}\,,\;\dot{I_{f}}\,,\;\dot{\phi_{f}}\;$ with $\;{\partial
h_f}/{\partial t}\;,\;{\partial I_f}/{\partial t}\;,\;{\partial
{\phi}_f}/{\partial t}\;$, because so far we have been considering the
situation of no disturbances turned on (i.e., the case when the full
derivatives coincide with the partial ones, and lack convective terms). Our
next step will be to turn on the disturbance $\,\Delta {\cal H}\,$, which will
include a transition from an inertial frame to the precessing frame of the
Earth's orbit. In accordance with formulae (\ref{14}) and (\ref{16}), this
transition will generate additions to the derivatives (\ref{24} - \ref{26}),
the additions that make the difference between a total and a partial
derivative.

~\\

 \noindent
 {\large{\textbf{A$\,$2.2 ~~Turning on the perturbation -- switching
 to a precessing frame}}}\\

 Our goal here is to derive the convective terms
 $\Phi_n=(\,\Phi_{_{h_f}}\,,\,\Phi_{_{I_f}}\,,\,\Phi_{_{{\phi}_f}}\,)$ that are
 to be added to the partial derivatives (\ref{24} - \ref{26}), to get the full
 time derivatives $\,\dot{q}_n=(\,\dot{h}_f\,,\,\dot{I}_f\,,\,\dot{\phi}_f\,)$.\\

 \noindent
 {{\textbf{A$\,$2.2.1 ~~Generalities}}}\\

As explained by
 Kinoshita (1977), the undisturbed dependence of the Euler angles of the Earth's figure
 upon the Andoyer elements can be approximated with
 \ba
 h_f\;=\;h\;+\;\frac{J}{\sin I}\;\sin
 g\;+\;O(J^2)\;\;\;,\;\;\;\;\;\;\;\;\;\;\;\;
 \label{27}
 \label{962}
 \label{460}
 \ea
 \ba
 I_f\;=\;I\;+\;J\;\cos
 g\;+\;O(J^2)\;\;\;,\;\;\;\;\;\;\;\;\;\;\;\;\;\;\;\;
 \label{28}
 \label{963}
 \label{461}
 \ea
 \ba
 \phi_f\;=\;{\it l}\;+\;g\;-\;J\;\cot I\;\sin g\;+\;O(J^2)\;\;\;,
 \label{29}
 \label{462}
 \label{964}
 \ea
 $J~$ and $~I~$ being the angles that the invariable plane (the one orthogonal to the
 angular momentum $~\bf\vec G~$) makes with the body equator and with the ecliptic plane
 of date, correspondingly. (For the Earth, $~J~$ is of order $~10^{-6}\,.\,$) As evident
 from Fig. 3, these angles are interconnected with the Andoyer variables $~L~$ and $~G~$
 through formulae
 \ba
 L\;=\;G\;\cos J\;\;\;
 \label{30}
 \label{463}
 \label{965}
 \ea
 and
 \ba
 H\;=\;G\;\cos I\;\;\;.
 \label{31}
 \label{966}
 \label{464}
 \ea
Under perturbations, formulae (\ref{27} - \ref{29}) will stay valid.
However, the expressions for the angles' evolution rate, (\ref{24} -
\ref{26}), will acquire convective additions (\ref{20}) caused by
the loss of osculation. These additions, entering the expressions
for
$\;\dot{q}_n\,=\,(\,\dot{h}_f\,,\;\dot{I}_f\,,\;\dot{\phi}_f\,)\,$,
will read, accordingly, as
 \ba
 {\Phi}_{h_f}\;=\;\frac{\partial \Delta \cal H}{\partial
 p_{h_f}}\;\;\;,\;\;\;\;
 {\Phi}_{I_f}\;=\;\frac{\partial \Delta \cal H}{\partial
 p_{I_f}}\;\;\;,\;\;\;\;
 {\Phi}_{\phi_f}\;=\;\frac{\partial \Delta \cal H}{\partial p_{\phi_f}}\;\;\;.
 \label{32}
 \label{465}
 \label{967}
 \ea
So our next step will be to calculate these three terms.

 Among the perturbations entering the Kinoshita theory, there is a so-called ``E term."
 It emerges due to a transition from an inertial frame to a noninertial one, i.e., from
 a coordinate system associated with the ecliptic of epoch to the one associated with the
 ecliptic of date. Simply speaking, in the Kinoshita theory the Earth rotation is
 considered in a noninertial frame of the terrestrial orbit precessing about the Sun.
 In Kinoshita (1977), the $\,xy\,$ plane of this noninertial frame is referred to as
 the {\emph{moving plane}}. In his theory, this ``E term" is the only one dependent not
 only upon the instantaneous orientation but also upon the angular velocity of the Earth
 (or, in the Hamiltonian formulation, upon the momenta conjugate to the Euler angles of
 the Earth's figure). Hence, in this situation $\;{\partial \Delta {\cal H}}/{\partial
 p_j}\,=\,{\partial E}/{\partial p_j}\;$. The expressions for $\,\Delta {\cal H}\,$ and the ``E term" are
 rendered by formulae (\ref{437} - \ref{438})
 where $\,p_{_{h_f}}\,,\,p_{_{I_f}}\,,\,p_{_{\phi_f}}\,$ denote the canonical momenta,
 while $\,\mu_x\,$, $\,\mu_y\,$, $\,\mu_z\,$ signify the body-frame components of the
 angular rate at which the orbit plane is precessing relative to an inertial coordinate
 system.\footnote{~We would
 point out that in Kinoshita's theory the origin of both $\,\Pi_1\,$ and $\,h\,$ is
 the mean equinox, whereas in our formalism the origin simply coincides with the $\,x
 \,$ axis. For our $\,\Pi_1\,$ and $\,h\,$ to coincide with those of Kinoshita, not
 only must we choose our inertial coordinate system with its $\,xy\,$ plane being
 within the ecliptic of epoch, but we should also choose the $\,x\,$ axis to coincide
 with the mean equinox of epoch. Similarly, not only should our precessing frame to be
 associated with the ecliptic of date, but the precessing $\,x\,$ axis whence we
 reckon the angles should be placed exactly at the angular distance of $\,-\,\Pi_1\,$
 from the node -- see Fig. 2 in Kinoshita (1977). (Mind that in the presence of
 precession Kinoshita employs notation $\,h'\,$ instead of $\,h\,$.)}

 In order to continue, we need the expressions for the body-frame components
 $\,\mu_x\,,\,\mu_y\,,\,\mu_z\,$. These can be obtained from the precessing-frame
 components $\,\mu_1\,,\,\mu_2\,,\,\mu_3\,$ by means of the appropriate rotation
 matrix:
 \ba
 \left[
 \begin{array}{ccc}
 \mu_x\\
 \mu_y\\
 \mu_z
 \end{array}
 \right]
 = ~~~~~~~~~~~~~~~~~~~~~~~~~~~~~~~~~~~~~~~~~~~~~~~~~~~~~~~~~~~~~~~~~~
   ~~~~~~~~~~~~~~~~~~~~~~~~~~~~~~~~~~~~~~~~~~~~~~~~~~~~
   \label{467}
   \label{969}
 \ea
 \ba
 \nonumber
 \left[
 \begin{array}{ccc}
 ~~\cos\phi_f\,\cos h_f\,-\,\sin\phi_f\,\cos I_f\,\sin h_f &
 ~~~\cos\phi_f\,\sin h_f\,+\,\sin\phi_f\,\cos I_f\,\cos h_f &
 ~\sin\phi_f\,\sin I_f \\
 -\sin\phi_f\,\cos h_f\,-\,\cos\phi_f\,\cos I_f\,\sin h_f &
 \,-\sin\phi_f\,\sin h_f\,+\,\cos\phi_f\,\cos I_f\,\cos h_f &
 ~\cos\phi_f\,\sin I_f \\
   \sin I_f\,\sin h_f ~~~~~~~~~~\,~~~~~~~~~~~~~~~~&
 -\sin I_f\,\cos h_f ~~~~~~~~~~~~~~~~~~~~~~~~~~~~ &
 ~\cos I_f~~~~~~~~ \\
 \end{array}
 \right]
 \left[
 \begin{array}{ccc}
 \mu_1\\
 \mu_2\\
 \mu_3
 \end{array}
 \right]~~~
 \label{}
 \ea
 Since no other contributions in $\;\Delta {\cal H}\;$ other than
 $\;E\;$ depend upon the momenta, then
 \ba
 {\Phi}_{h_f}\,=\,\frac{\partial\Delta\cal H}{\partial
 p_{_{h_f}}}\,=\,\frac{\partial E}{\partial p_{_{h_f}}}
 \,=\,-\;\frac{\sin \phi_f}{\sin I_f}\;\mu_x\,-\;\frac{\cos \phi_f}{\sin
 I_f}\;\mu_y
  =\,{\mu}_1\,\cot I_f\,\sin h_f\,-\,{\mu}_2\,\cot I_f\,\cos h_f\,-\,{\mu}_3
 ~~~,~\,~~
 \label{35}
 \label{468}
 \ea
 \ba
 {\Phi}_{I_f}\;=\;\frac{\partial \Delta {\cal H} }{\partial p_{_{I_f}}}\;=\;
 \frac{\partial E}{\partial
 p_{_{I_f}}}\;=\;-\;\mu_x\;\cos \phi_f\;+\;\mu_y\;\sin \phi_f\,=\;
 -\;{\mu}_1\;\cos h_f\;-\;{\mu}_2\;\sin h_f~~~,~~~~~~~~~~~~~~~~~~~~\,
 \label{36}
 \label{469}
 \ea
 \ba
 {\Phi}_{\phi_f}\,=\;\frac{\partial \Delta {\cal H} }{\partial p_{_{\phi_f}}}\,=\;
 \frac{\partial E}{\partial
 p_{_{\phi_f}}}\,=\,\frac{\sin \phi_f\;\cos I_f}{\sin
 I_f}\,\mu_x\,+\,
 \frac{\cos \phi_f\;\cos I_f}{\sin I_f}\,\mu_y\,-\,\mu_z
 =\;-\;\frac{\sin h_f}{\sin I_f}\;{\mu}_1\,+\;
 \frac{\cos h_f}{\sin I_f}\;{\mu}_2\;\;\,.~~~
 \label{37}
 \label{470}
 \ea
 Naturally, none of the $\;\Phi\;$ terms bears dependence upon
 $\;\phi_{\textstyle{_f}}\;$.

 To write down the $\;\Phi\;$ terms as functions of the longitude $\;\Pi_1\,$ and
 inclination $\;\pi_1\,$ of the ecliptic of date on that of epoch, we shall insert
 into (\ref{468} - \ref{470}) the appropriate expressions for $\,\mu_1\,,\;\mu_2\,,
 \;\mu_3\,$. However, at this point care is needed, because of the freedom of choice
 of a coordinate system co-precessing with the orbital plane.\footnote{~We are  grateful to Hiroshi Kinoshita who explained to us the choice accepted in his works.}


 \ba
 \nonumber
 \mbox{\bf{\textbf{A$\,$2.2.2 ~~The precession rate $\;\mubold\;$ as seen in a certain
 coordinate system}}}~~~~~~~~~~\\
 \nonumber
 \mbox{\bf{\textbf{associated with the precessing equator of date}}}
 ~~~~~~~~~~~~~~~~~~~~~~~~~~~~~~~~~
 \ea

 Let the inertial axes $\;(\,X\,,\;Y\,,\;Z\,)\;$ be fixed in
 space so that $\;X\;$ and $\;Y\;$ belong to the ecliptic of epoch. A rotation within
 the ecliptic-of-epoch plane by longitude $\;\Pi_1\;$, from the axis

 \pagebreak

 \noindent
 $X\;$, will define
 the line of nodes. A rotation about this line by an inclination angle
 $\;\pi_1\;$ will give us the ecliptic of date. The line of nodes, $\;1\;$, along with
 axis $\;2\;$ naturally chosen within the ecliptic-of-date plane, and with axis
 $\;3\;$ orthogonal to this plane, will constitute the precessing coordinate system,
 with the appropriate basis denoted by $\;(\,\mathbf{\hat{e}}_1\,,\;\mathbf{\hat{e}}_2
 \,,\;\mathbf{\hat{e}}_3\,)\;$. For example, the unit vector $\,\mathbf{\hat{e}}_3\,$
 reads in the inertial axes $\,(\,{{X}}\,,\;{{Y}}\,,\;{{Z}}\,)\,$ as
 \ba
 \mathbf{\hat{e}}_3\;=\;\left(\;\sin \pi_1\,\sin \Pi_1\;\;,\;\;
 \;-\,\sin \pi_1\,\cos\Pi_1\;\;,\;\,\;\cos\pi_1\;\right)^{^T}\;\;\;.
 \label{z}
 \ea
 The Earth's angular velocity relative to the inertial and precessing axes
 obey
 \ba
 \omegabold^{(inert)}\;=\;\omegabold^{(rel)}\;+\;\mubold\;\;\;,
 \label{}
 \ea
 $\mubold\,$ being the precession rate of the precessing axes $\,{\bf{\hat{e}}}_j\,$
 relative to the inertial axes $\,(\,{{X}}\,,\;{{Y}}\,,\;{{Z}}\,)\,$. In the inertial
 axes, this rate is given by
 \ba
 \mubold{^{\,}}'\;=\;\left(\;
 {\dot{\pi}}_1\,\cos \Pi_1\;\;\;,\;\;\;\,{\dot{\pi}}_1\,\sin \Pi_1
  \;\;\;,\;\;\,\;{\dot{\Pi}}_1\;\right)^{^T}\;\;\;,
 \label{muinertial}
 \ea
 because this expression satisfies the equality $\;\mubold{^{\,}}'\;\times\;
 {\bf{\hat{e}}}_3\;=\; {\bf{\dot{\hat{e}}}}_3\;$, as can be easily seen from
(\ref{z}) and (\ref{muinertial}).

 In a frame precessing with the ecliptic, the precession rate will be represented by
 the vector
 \ba
 \mubold\;=\;{\bf{\hat{R}}}_{e\rightarrow d}\;\;\mubold{^{\,}}'\;\;\;,
 \label{baba}
 \ea
 where
   \ba
  {\bf{\hat{R}}}_{e\rightarrow d}=\,{\bf{\hat{R}}}_{1}(\pi_1)\,
 {\bf{\hat{R}}}_{Z}(\Pi_1)=
 \left[
 \begin{array}{ccc}
    ~~~~~~~~~~~~\, \cos \Pi_1   &    ~~~~~~~~~~~~~~~~\,\sin \Pi_1          &  ~~~~~~~~~~~~ 0 \\
\\
  \, -\;\cos \pi_1 \;\sin \Pi_1 &  \;\;\;\;\;\;\;\;\;\,\cos \pi_1\;\,\cos \Pi_1  &   ~~~~~~~~~\sin \pi_1  \\
\\
  ~~~~~~\;\sin \pi_1\;\,\sin \Pi_1\;\;\;  & \;\;\;\;\;-\;\sin \pi_1\,\;\cos \Pi_1  &  ~~~~~~~~~\cos \pi_1
\end{array}
 \right]~~~~
 \label{ded}
 \ea
 is the matrix of rotation from the ecliptic of epoch to that of date. From
 (\ref{baba} - \ref{ded}) we get the
 components of the precession rate,\footnote{~Equivalently, one can
 find the components of $\,\mubold\,$ as the elements of the
 skew-symmetric matrix $\,{\bf{\dot{\hat{R}}}}_{e\rightarrow d}
 \,{\bf{\hat{R}}}_{e\rightarrow d}^{\textstyle{^{-1}}}\,$.} as seen in the
 co-precessing coordinate frame $\;(1\,,\;2\,,\;3)\;$:
 \ba
 \mubold\;=\;\left(\;\mu_1\;,\;\;\mu_2\;,\;\;\mu_3\;\right)^{^T}\;=\;\left(\;
 {\dot{\pi}}_1\;\;,\;\;\;\,{\dot{\Pi}}_1\,\sin \pi_1
  \;\;,\,\;\;\;{\dot{\Pi}}_1\,\cos \pi_1\;\right)^{^T}\;\;\;.
   \label{34}
  \label{466}
 \ea
 Substitution of these components into (\ref{468} - \ref{470}) entails:
 \ba
 {\Phi}_{h_f}\,=\,\dot{\pi}_1\,\cot I_f\,\sin h_f\,-\,\dot{\Pi}_1\,\sin
 \pi_1\,\cot I_f\,\cos h_f\,-\,\dot{\Pi}_1\,\cos\pi_1 ~~~,~~~~~~~~~~~
 \label{970}
 \ea
 \ba
 {\Phi}_{I_f}\;=\;-\;\dot{\pi}_1\;\cos h_f\;-\;\dot{\Pi}_1\;\sin\pi_1\;
 \sin h_f~~~,~~~~~~~~~~~~~~~~~~~~~~~~~~~~~~~~~~~~~~~
 \label{971}
 \ea
 \ba
 {\Phi}_{\phi_f}\;=\;-\;\frac{\sin h_f}{\sin I_f}\;\dot{\pi}_1\;+\;
 \frac{\cos h_f}{\sin I_f}\;\dot{\Pi}_1\;\sin\pi_1\;\;\;.~~~~~~
 ~~~~~~~~~~~~~~~~~~~~~~~~~~~~~~~~~
 \label{972}
 \ea

 \ba
 \nonumber
 \mbox{\bf{\textbf{A$\,$2.2.3 ~~The precession rate $\;\mubold\;$ as seen in a
 different coordinate system}}}~~~~~~~~~~\\
 \nonumber
 \mbox{\bf{\textbf{associated with the precessing equator of date}}}
 ~~~~~~~~~~~~~~~~~~~~~~~~~~~~~~~~~~~\\
 \nonumber
 \mbox{\bf{\textbf{(the system used by Kinoshita
 1977)}}}~~~~~~~~~~~~~~~
 ~~~~~~~~~~~~~~~~~~~~~~~~~~~~~~~~~~~
 \ea

 In the preceding subsection the transition from the ecliptic of epoch to the
 one of date was implemented by two Euler rotations: $\;{\bf{\hat{R}}}_{e
 \rightarrow d}\;=\;{\bf{\hat{R}}}_{N}(\pi_1)\;{\bf{\hat{R}}}_{Z}(\Pi_1)\;=
 \;{\bf{\hat{R}}}_{1}(\pi_1)\;{\bf{\hat{R}}}_{Z}(\Pi_1)\;$.
 The axis $\,1\,$ of the precessing frame was assumed to
 coincide with the line of nodes, $\,N\,$. Evidently, this choice was just
 one out of an
 infinite multitude. An alternative option was employed by Kinoshita (1977),
 who used a sequence of three Eulerian rotations: $\;{\bf{\hat{R}}}_{e
 \rightarrow d}^{K}\;=\;{\bf{\hat{R}}}_{3}(\,-\,\Pi_1)\;{\bf{\hat{R}}}_{N}
 (\pi_1)\;{\bf{\hat{R}}}_{Z}(\Pi_1)\;$. Specifically, having performed
 the two rotations described above, Kinoshita then rotated the axis $\,1\,$ within
 the ecliptic of date by angle $\;-\,\Pi_1\,$ away from the line of nodes $\,N\,$.
 Due to reasoning analogous to what was presented in the subsection above, the
 sequence of three rotations gives, instead of (\ref{466}), the following
 expression:\footnote{~In the precessing frame, the angular momentum reads:
 $\,(\,G\;\sin I\;\sin h\,,\;
 -\;G\;\sin I\;\cos h\,,\;H)^{\textstyle{^T}}$, quantities $\,I\,$,
 $\,h\,$, and $\,H\,$ being as in Fig. 3. This, together with (\ref{968})
 and the formula
 $\,\Delta{\cal H}\,=\,E\,=\,-\,
 \mubold\cdot{\vec{\bf{G}}}\,$,  will entail:
 \ba
 \nonumber
 \Delta{\cal H}\,=\,E\,=\;\dot{\Pi}_1\;H\;\left(1\,-\,\cos \pi_1 \right)\;
 -\;\dot{\pi}_1\;G\;\sin I\;\sin(h\,-\,\Pi_1)\;+\;\dot{\Pi}_1\;G\;\sin I\;
 \cos(h\,-\,\Pi_1)\;\sin \pi_1\;\;\;,
 \ea
 which coincides with expression (3.4) in Kinoshita (1977).}
 \ba
 \nonumber
 \mubold\;=\;\left(\;\mu_1\;,\;\;\mu_2\;,\;\;\mu_3\;\right)^{^T}
 ~~~~~~~~~~~~~~~~~~~~
 ~~~~~~~~~~~~~~~~~~~~~~~~~~~~~~~~~~~~~~~~~~~~~~~~~~~~~~~~~~~~~~~~~~~~~~~~~~~~~~~~\\
 \label{968}\\
 \nonumber
 =\left(\,{\dot{\pi}}_1\;\cos\Pi_1\,-\;\dot{\Pi}\;\sin\Pi_1\;\sin\pi_1
 \;\;,\;\;\;\,\dot{\pi}_1\;\sin\Pi_1\,+\;{\dot{\Pi}}_1\;\cos\Pi_1\;\sin \pi_1
 \;\;,\,\;\;\;{\dot{\Pi}}_1\,\cos \pi_1\,-\;\dot{\Pi}_1\,\right)^{^T}\;\;\;.
 ~~~~~\,
 \ea
 Insertion thereof into (\ref{468} - \ref{470}) will yield:
 \ba
 {\Phi}_{h_f}\,=\,\dot{\pi}_1\,\cot I_f\,\sin (h_f\,-\,\Pi_1)\,-\,\dot{\Pi}_1\,\sin
 \pi_1\,\cot I_f\,\cos (h_f\,-\,\Pi_1)\,-\,\dot{\Pi}_1\,\cos\pi_1\,+\,\Pi_1 ~~~,~~~
 \label{970bis}
 \ea
 \ba
 {\Phi}_{I_f}\;=\;-\;\dot{\pi}_1\;\cos (h_f\,-\;\Pi_1)\;-\;\dot{\Pi}_1\;\sin\pi_1\;
 \sin (h_f\,-\;\Pi_1)~~~,~~~~~~~~~~~~~~~~~~~~~~~~~~~~~~~~~~~~~~~
 \label{971bis}
 \ea
 \ba
 {\Phi}_{\phi_f}\;=\;-\;\frac{\sin (h_f\,-\;\Pi_1)}{\sin I_f}\;\dot{\pi}_1\;+\;
 \frac{\cos (h_f\,-\;\Pi_1)}{\sin I_f}\;\dot{\Pi}_1\;\sin\pi_1\;\;\;.~~~~~~
 ~~~~~~~~~~~~~~~~~~~~~~~~~~~~~~~~~
 \label{972bis}
 \ea\\

 \noindent
 {\large{\textbf{A$\,$2.3 ~~The perturbed velocities}}}\\

 According to (\ref{18}), to get the full evolution of the
 figure-axis Euler angles under perturbation, one should sum the
 unperturbed velocities, given by the partial derivatives
 (\ref{24} - \ref{26}),
  with the appropriate convective terms (\ref{35} - \ref{37}):
 \ba
 \dot{h}_f\;=\;\left(\frac{\partial h_f}{\partial t}\right)_C\,+\;\Phi_{_{h_f}}
 ~=
~~~~~~~~~~~~~~~~~~~~~~~~~~~~~~~~~~~~~~~~~~~~~~~~~~~~~~~~~~~~~~~~~~~~~~~~~~~
~~~~~~~~~~~~~~~~~~~~~
 \label{38}
 \label{471}
 \label{973}
 \ea
 \ba
 \nonumber
 \frac{1}{\sin I_{f}}\,\sqrt{G^2\,-\,L^2}\,\left[ \frac{\sin {\it l}\,\sin\phi_{f}}{A}\,+
 \,\frac{\cos {\it l}\,\cos\phi_{f}}{B} \right]\,+\,\dot{\pi}_1\,\cot I_f\,\sin h_f\,-\,
 \dot{\Pi}_1\,\sin\pi_1\,
 \cot I_f\,\cos h_f\,-\,\dot{\Pi}_1\,\cos\pi_1 ~~,~~
 \ea
 \pagebreak
 \ba
 \nonumber
 \dot{I}_f\;=\;\left(\frac{\partial I_f}{\partial t}\right)_C\,+\;\Phi_{_{I_f}}
 ~~~~~~~~~~~~~~~~~~~~~~~~~~~~~~~~~~~~~~~~~~~~~~~~~~~~~~~~~~~~~~~~~~~~~~~~~~~~~~~~~~
 ~~~~~~~~~~~~~~~~~~~
 \ea
 \ba
 =\;\sqrt{G^2\,-\,L^2}\;
  \left[ \frac{\sin {\it l}\;\cos\phi_{f}}{A}\;-\;\frac{\cos {\it l}\;\sin\phi_{f}}{B}
 \right]~-\;\dot{\pi}_1\;\cos h_f\;-\;\dot{\Pi}_1\;\sin\pi_1\;\sin
 h_f~~~,~~~~~~~~~
 \label{39}
  \label{472}
  \label{974}
 \ea
 \ba
 \nonumber
 \dot{\phi}_f\;=\;\left(\frac{\partial \phi_f}{\partial
 t}\right)_C\,+\;\Phi_{_{\phi_f}}~~~~~~~~~~~~~~~~~~~~~~~~~~~~~~~~~
 ~~~~~~~~~~~~~~~~~~~~~~~~~~~~~~~~~~~~~~~~~~~~~~~~~~~~~~~~~~~
 \ea
 \ba
 =\;\frac{L}{C}\,-\,\sqrt{G^2\,-\,L^2}\,\cot I_{f}\,
 \left[ \frac{\sin {\it l}\,\sin\phi_{f}}{A}\,+\,\frac{\cos {\it l}\,\cos\phi_{f}}{B}
 \right]~-\;\frac{\sin h_f}{\sin I_f}\;\dot{\pi}_1\;+\;
 \frac{\cos h_f}{\sin I_f}\;\dot{\Pi}_1\;\sin\pi_1
 \;\;.~~~
 \label{40}
  \label{473}
  \label{975}
 \ea
 These expressions will help
 us to determine the instantaneous orientation of the spin
 axis.

  ~\\


  \noindent
 {\large{\textbf{A$\,$2.4 ~~The precessing-frame-related angular velocity}}}\\
 $\left.~~~~~~~~~~~~\right.$ {\large{\textbf{expressed through the Andoyer elements}}}\\
 $\left.~~~~~~~~~~~~\right.$ {\large{\textbf{introduced in the precessing frame.}}}\\

 Let angles $\,h_r^{^{(rel)}}\,$ and $\,I_r^{^{(rel)}}\,$ be the precessing-frame-related
 node and inclination of vector of the angular velocity relative to the precessing frame,
 and let $\;\omega^{^{(rel)}}\,\equiv\,|\,\omegabold^{(rel)}\,|\;$ be the relative spin
 rate. Our next step will be to derive expressions for $\,h_r^{^{(rel)}}\,$ and
 $\,I_r^{^{(rel)}}\,$ as functions of $\,h_f\,,\,I_f\,,\,\phi_f\,,\,\dot{h}_f\,,\,
 \dot{I}_f\,,\,\dot{\phi}_f\,$, and then to substitute the expressions (\ref{38} -
 \ref{40}) for $\;\dot{h}_f\,,\,\dot{I}_f\,,\,\dot{\phi}_f\;$ therein, in order to express
 $\,h_r^{^{(rel)}}\,$ and $\,I_r^{^{(rel)}}\,$ via $\,h_f\,,\,I_f\,,\,\phi_f\,$ only. To
 accomplish this step, let us begin with the formulae interconnecting the precessing-frame
 components of the angular velocity relative to the precessing frame with the figure-axis
 Euler angles and with the spin-axis Euler angles. These formulae are fundamental and
 perturbation-invariant:
 \ba
 \omega^{(rel)}_1\;=\;\dot{I}_f\;\cos\,h_f\;+\;\dot{\phi}_f\;\sin
 I_f\;\sin\,h_f\,
 \label{41}
 \label{474}
 \label{976}\\
 \nonumber\\
 \omega^{(rel)}_2\;=\;\dot{I}_f\;\sin\,h_f\;-\;\dot{\phi}_f\;\sin \,I_f\;\cos\,h_f
 \label{42}
 \label{977}
 \label{475}\\
 \nonumber\\
 \omega^{(rel)}_3\;=\;\dot{h}_f\;+\;\dot{\phi}_f\;\cos I_f
 ~~~~~~~~~~~~~~~~~~\,
 \label{43}
 \label{476}
 \label{978}
 \ea
and
 \ba
 \omega^{(rel)}_1\;=\;\omega^{^{(rel)}}\;\sin \,I_r^{^{(rel)}}\;\sin \,h_r^{^{(rel)}} ~~~~
 \label{44}
 \label{477}\\
 \label{979}
 \nonumber\\
 \omega^{(rel)}_2\;=\;-\;\omega^{^{(rel)}}\;\sin \,I_r^{^{(rel)}}\;\cos \,h_r^{^{(rel)}}\,
 \label{45}
 \label{478}
 \label{980}\\
 \nonumber\\
 \omega^{(rel)}_3\;=\;\omega^{^{(rel)}}\;\cos I_r^{^{(rel)}} ~~~.~~~~~~~~~~~~
 \label{46}
 \label{479}
 \label{981}
 \ea
 Both the inertial and relative spin rates, $\,\omega^{^{(inert)}}\,$ and $\,
 \omega^{^{(rel)}}\,$, can be most conveniently calculated in the body frame. In
 that frame, the inertial angular velocity can be written as
 \ba
 \omegabold^{^{(inert)}}\;=\;\left(\;\frac{L_x}{A}\;,\;\;\;
 \frac{L_y}{B}\;,\;\;\;\frac{L_z}{C}\;\right)^{\textstyle{^{T}}}=\;
 \left(\;\frac{G\;\sin J\;\sin {\it{l}}}{A}\;\;,\;\;\;\;\frac{G\;\sin J\;\cos {\it{l}}}{B}
 \;\;,\;\;\;\;\frac{G\;\cos J}{C}\;\right)^{\textstyle{^{T}}}\;\;\;,\;\;\;\;\;
 \label{inertial}
 \label{982}
 \ea
 whence its absolute value turns out to be
 \ba
 \nonumber
 \omega^{^{(inert)}}\;
 =\;
 \sqrt{\;\left(\frac{L_x}{A}\right)^2\;+\;
         \left(\frac{L_y}{B}\right)^2\;+\;
         \left(\frac{L_z}{C}\right)^2\;
 }\;=~~~~~~~~~~~~~~~~~~~~~~~~~~~~~~~~~~~\\
 \nonumber\\
 \nonumber\\
 \sqrt{\left(\frac{G\;\sin J\;\sin \it l}{A}\right)^2 +\,
         \left(\frac{G\;\sin J\;\cos \it l}{B}\right)^2 +\,
         \left(\frac{G\;\cos J}{C}\right)^2
 }
 =\,\frac{G}{C} \left[1 + O\left(J^2\right)  \right] =
  \,\frac{L}{C} \left[1 + O\left(J^2\right)  \right]\;\;.\;\;\;
 \label{47}
 \label{480}
 \label{983}
 \ea
 To derive the relative rate, square the obvious equality
 $\,\omegabold^{^{(rel)}}\,=\,\omegabold^{^{(inert)}}-\,\mubold\;$, to obtain:
 \ba
 (\,\omegabold^{^{(rel)}}\,)^{{\textstyle{^2}}}=\;
 (\,\omegabold^{^{(inert)}}\,)^{{\textstyle{^2}}}
 -\;2\;\omegabold^{^{(inert)}}\cdot\;\mubold\;+\;\mubold^{\textstyle{~^2}}\;\;\;.
 \label{984}
 \ea
 Hence,
 \ba
 \omega^{^{(rel)}}\;=\;\omega^{^{(inert)}}\left[\;1\;-\;\alpha\;+\;O\left(\,\left(
 {{\Phi}}/{\omega}\,\right)^2\,\right)\;\right]\;=\;\frac{L}{C}\;\left[\;1\;-\;\alpha\;+\;
 O\left(\,\left( {{\Phi}}/{\omega}\,\right)^2\;+\;O(J^2)\,\right)\;\right]
 \;\;\;\;\;\;\;\;\;\;\;
 \label{985}
 \ea
 and
 \ba
 \frac{1}{\omega^{^{(rel)}}}\;=\;\frac{1}{\omega^{^{(inert)}}}\left[\;1\;+\;\alpha\;+\;
 O\left(\,\left({{\Phi}}/{\omega}\,\right)^2\,\right)\;\right]\;=\;\frac{C}{L}\;\left[\;1
 \;+\;\alpha\;+\;O\left(\,\left({{\Phi}}/{\omega}\,\right)^2\,\right)\;+\;O(J^2)\,\right]
 \;\;,\;\;\;\;\;\;
 \label{986}
 \ea
 where
 \ba
 \alpha\;\equiv\;\frac{\omegabold^{^{(inert)}}\cdot
 \;\mubold}{(\,\omegabold^{^{(inert)}}\,)^{\textstyle{^2}}}
 \;\;\;.
 \label{987}
 \ea
 is of order $\,\Phi/\omega\,$. Dot-multiplying (\ref{982}) by (\ref{969}), we arrive at:
 \ba
 \nonumber
 \alpha\,=\frac{G~\cos J}{C}~\frac{\mu_z~~~}{~~(\,\omegabold^{^{(inert)}}\,)^{\textstyle{^2}}}\,+\,
 O(J\Phi/\omega)~~~~~~~~~~~~~~~~~~~~~~~~~~~~~~~~~~~~~~~~~~~~~~~~~~~~~~~~~~
 ~~~~~~~~~~~~~~~~~~~~\\
 \nonumber\\
 =\,\frac{C}{L}\;\left[\,\mu_1\,\sin I_f\,\sin h_f
 \,-\,\mu_2\,\sin I_f\,\cos h_f\,+\,\mu_3\,\cos I_f\,\right]\,+\,O(J\Phi/\omega)
 \,+\,O(J^2)~~~.~~~~~~~~~~~~~~~~~~~~
 \label{alpha}
 \ea
 In the coordinate system described in A.2.2.2, substitution of
 (\ref{34}) makes the latter read
 \ba
 \alpha\,=\,\frac{C}{L}\,\left(\,{\dot{\pi}}_1\,\sin I_f\,\sin h_f\,-\,{\dot{\Pi}}_1
 \,\sin\pi_1\,\sin I_f\,\cos h_f\,+\,{\dot{\Pi}}_1\,\cos\pi_1\,\cos I_f\right)\,+
 \,O(J\Phi/\omega)\,+\,O(J^2)~~.
 ~~~\,
 \label{988}
 \ea
 In Kinoshita's coordinate system described in subsection A.2.2.3, one should use for
 the components of $\,\mubold\,$ not expression (\ref{34}) but (\ref{968}), insertion
 whereof into (\ref{alpha}) results in:
 \ba
 \nonumber
 \alpha\,=~\frac{C}{L}\,\left(\,{\dot{\pi}}_1\,\sin I_f\,\sin (h_f\,-\,\Pi_1)\,-\,
 {\dot{\Pi}}_1\,\sin\pi_1\,\sin I_f\,\cos (h_f\,-\,\Pi_1)\,+\,{\dot{\Pi}}_1\,\cos\pi_1
 \,\cos I_f\right.\,~~~~~~~~~~~~~\\
  \label{1988}\\
  \nonumber
 \left.-\;\,\dot{\Pi}_1\,\cos I_f\,\right)\,+\,O(J\Phi/\omega)\,+\,O(J^2)~~.
 ~~~~~~~~~~~~~~~~~~~~~~~~~~~~~~~~~~~~~~~~~~~~~~~~~~~~~~~~~~~~~~~~~~~~~~\;
 \ea
 This formula will enable us to derive approximate (valid up to $\;O\left(J^2\right)\,+\,O
 \left(J\Phi/\omega\right)\,+\,O\left((\Phi/\omega)^2\right)\;\;$) expressions for $\;h_r^{^{
 (rel)}}\;$ and $\;I_r^{^{(rel)}}\;$ expressed as functions of the Andoyer
 variables.

 \noindent
{\underline{\bf{\Large{\textbf{Appendix 3.~~~Expression for
$\;I_r^{^{(rel)}}\;$}}}}}
 ~\\

  ~\\
Together, (\ref{43}) and (\ref{46}) will give:
 \ba
 \omega^{^{(rel)}}\;\cos I_r^{^{(rel)}}\;=\;\dot{h}_f\;+\;\dot{\phi}_f\;\cos I_f
 \label{A11}
 \label{481}
 \label{989}
 \ea
or, equivalently,
 \ba
 \omega^{^{(rel)}}\;\left(\;\cos I_r^{^{(rel)}}\;-\;\cos I_f\;\right)\;=\;
 \dot{h}_f\;+\;\left(\dot{\phi}_f\;-\;\omega^{^{(rel)}}\;\right)\;\cos I_f\;\;.
 \label{A12}
 \label{482}
 \label{990}
 \ea
Since we are planning to carry out all calculations neglecting terms
$\;O\left(J^2\right)\;$, and since the three inclinations
$\;I_f\,,\,I_r^{^{(rel)}}\,,\,I\;$ differ from one another by
quantities of order $\;O(J)\;$, we can approximate the left-hand
side of (\ref{A12}) with the first-order terms of its Taylor
expansion:
 \ba
 \omega^{^{(rel)}}\;\left(\;-\;\sin \,I_f\;\right)\;
 \left(\;I_r^{^{(rel)}}\;-\;I_f\;\right)\;=\;
 \dot{h}_f\;+\;\left(\dot{\phi}_f\;-\;\omega^{^{(rel)}}\;\right)\;
 \cos I_f\;+\;O\left(J^2\right)\;\;\;,
 \label{A13}
 \label{483}
 \label{991}
 \ea
wherefrom
 \ba
 \nonumber
 I_r^{^{(rel)}}\;-\;I_f\;=\;-\;
 \frac{\dot{h}_f}{\omega^{^{(rel)}}}\;\frac{1}{\sin \,I_f}\;+\;
 \left[\frac{\dot{\phi}_f}{\omega^{^{(rel)}}}\;-\;1\;\right]\;\left(\;-\;\cot I_f\right)
 \;+\;O\left(J^2\right)~~~~~~~~~~~~~~~~~~~~~~~~~~~~~~~~~~~~~~\\
 \nonumber\\
 \nonumber\\
 =\;-\;\left( \frac{\partial {h}_f}{\partial t}\,+\,\Phi_{h_f} \right)\;
 \frac{1}{\omega^{^{(rel)}}}\,\frac{1}{\sin \,I_f}\;+\;
 \left[ \left(\frac{\partial {\phi}_f}{\partial t}\,+\,\Phi_{\phi_f} \right)\,
 \frac{1}{\omega^{^{(rel)}}}\;-
 \;1\;\right]\;\left(\;-\;\cot I_f\right) \;+\;O\left(J^2\right)\;\;\;.\;\;\;\;\;\;~~~
 \label{A14}
 \label{484}
 \label{992}
 \ea
 To get rid of the time derivatives, employ formulae
 (\ref{38} - \ref{40}) or (\ref{24} - \ref{26}):
 \ba
 \nonumber
 I_r^{^{(rel)}}\;-\;I_f\;=\;\cot I_f\;\left\{\;1\;-\;\frac{1}{\omega^{^{(rel)}}}\;\left[
 \frac{L}{C}\;-\;\sqrt{G^2\,-\,L^2}\;\cot I_{f}\;
 \left( \frac{\sin {\it l}\;\sin\phi_{f}}{A}\;+\;\frac{\cos {\it l}\;\cos\phi_{f}}{B}
 \right)
 \right]\;\right\}\;- ~~\\
 \nonumber\\
 \label{A15}
 \label{485}
 \label{993}\\
 \nonumber
 \frac{1}{\omega^{^{(rel)}}\,\sin^2
 I_{f}}\,\sqrt{G^2-L^2}\,
 \left( \frac{\sin {\it l}\;\sin\phi_{f}}{A}\,+\,\frac{\cos {\it l}\;\cos\phi_{f}}{B}
 \right)-\,\frac{1}{\omega^{^{(rel)}}\,\sin \,I_f}\left(\Phi_{h_f}
 +\,\Phi_{\phi_f}\;\cos I_f
 \right)\,+\,O\left(J^2\right)
 \ea
 For $\,\phi_f\,$ we can use the approximation
 $\,\phi_f\,=\,{\it l}\,+\,\mbox{g}\,-\,J\,\cot I\,\sin\mbox{g}\,+\,O(J^2)\,$. Besides, in
 the terms of order $\,J\,$ and of order $\,\Phi/\omega\,$ we can substitute
 $\,\omega^{^{(rel)}}\,$ with $\,\omega^{^{(inert)}}=\,L/C\,+\,O(J^2)\,$. Such
 substitutions will entail errors of orders $\,O(J\Phi/\omega)\,$ and
 $\,O(\,(\Phi/\omega)^2\,)\,$. However, in the leading term we must use
 (\ref{985} - \ref{988}). Thus we get:
  \ba
 \nonumber
 I_r^{^{(rel)}}-\,I_f\,=\;\cot I_f\,\left(\,1\,-\,\frac{1}{\omega^{^{(rel)}}}\,\frac{L}{C}\,\right)\,+\,
 \frac{\sqrt{G^2\,-\,L^2}}{\omega^{^{(rel)}}}\,\left(\cot^2I_f\,-\,\frac{1}{\sin^2
 I_{f}}\right)\,\left[ \frac{\sin {\it l}\;\sin\phi_{f}}{A}\,+\,\frac{\cos {\it l}\;\cos\phi_{f}}{B}\,
 \right]\;\;\;\;\;\\
 \nonumber\\
 \nonumber\\
 \nonumber
 -\;\frac{1}{\omega^{^{(rel)}}}\;\frac{1}{\sin \,I_f}\;\left(\Phi_{h_f}
 +\,\Phi_{\phi_f}\;\cos I_f
 \right)\;+\;O(J^2)~=~~~~~~~~~~~~~~~~~~~~~~~~~~~~~~
 \ea
 ~\\
 \ba
 \nonumber
 \cot I_f \left(1
 -\frac{1\,+\,\alpha}{\omega^{^{(inert)}}}\;\,\frac{L}{C}\right)-\frac{G\,\sin J}{G/C}\left[
 \frac{\sin {\it l}}{A}\, \sin\left(
 {\it l}+\mbox{g}-J\,\cot I\,\sin \mbox{g}
 \right) +
 \frac{\cos {\it l}}{B}\,\cos\left(
 {\it l}+\mbox{g}-J\,\cot I\,\sin \mbox{g}
 \right)
 \right]
 \;\;\;\;\;\;\;\;
 \ea
 ~\\
 \ba
 \nonumber
  -\,\frac{1}{L/C}\,\frac{1}{\sin \,I_f}\;\left(\Phi_{h_f}
 +\,\Phi_{\phi_f}\;\cos I_f
 \right)\;+\;O\left(J^2\right)\;+\;O\left(J\,\Phi/\omega\right)\;
 +\;O(\,(\Phi/\omega)^2\,)~~~~~~\;\;\;\;\;\;\;\;\;
 \ea
 \ba
 \nonumber\\
 \nonumber
 =\;-\;\alpha\;\cot I_f\;-\;C\,J\,\left[\frac{\sin {\it l}}{A}\; \sin\left({\it l}\,+
 \,\mbox{g}\right)\,+\,\frac{\cos {\it l}}{B}\;\cos\left(
 {\it l}\,+\,\mbox{g}\right)\right]
 \ea
 \ba
 -\,\frac{C}{L}\,\frac{1}{\sin \,I_f}\,\left(\Phi_{h_f}+\,\Phi_{\phi_f}\;\cos I_f
 \right)
 \,+\,O\left(J^2\right)\,+\,O\left(J\Phi/\omega\right)\,+\,
 O\left(\,(\Phi/\omega)^2\,\right)~~,~~
 \label{A16}
 \label{486}
 \label{994}
 \ea
whence, by using (\ref{988}) and the formula
$\;I_f\,=\,I\,+\,J\,\cos g\,+\,O(J^2)\;$, we arrive at:
 \ba
 \nonumber
 I_r^{^{(rel)}}=I+J\left\{\cos \mbox{g}\,-\frac{C}{A}\;\sin {\it l}\;\sin\left(
 {\it l}+\mbox{g}
 \right)\,-\,\frac{C}{B}\;\cos {\it l}\;\cos\left(
 {\it l} + \mbox{g}
 \right)\right\}
   -\;
  \alpha\;\cot I_f ~~~~~
 \ea
 \ba
 \nonumber
 -\,\frac{C}{L}\,\frac{1}{\sin\,I_f}\,\left(\Phi_{h_f}
 + \Phi_{\phi_f}\;\cos I_f\right)\,+\,O\left(J^2\right)\,+\,O\left(J\Phi/\omega\right)\,+\,
 O\left(\,(\Phi/\omega)^2\,\right)
 \ea
 Via trigonometric transformations, the second term gets simplified as:
 \ba
 \cos \mbox{g}\,-\frac{C}{A}\;\sin {\it l}\;\sin\left({\it l}+\mbox{g}\right)\,-\,
 \frac{C}{B}\;\cos {\it l}\;\cos\left({\it l} + \mbox{g}\right)\;=\;\cos\mbox{g}\,-\,
 \frac{C}{A}\,\,\frac{\cos\mbox{g}\,-\;\cos(2\,{\it l}\,+\,\mbox{g})}{2} \\
 \nonumber\\
 \nonumber\\
 \nonumber
 -\,\frac{C}{B}\,\,\frac{\cos \mbox{g}\,+\;\cos(2\,{\it l}\,+\,\mbox{g})}{2}\;=\;\left(1\,
 -\,\frac{C}{2A}\,-\,\frac{C}{2B}\right)\,\left[\,\cos\mbox{g}\,-\,{\it e}\,
 \cos \left(2\,{\it l}+\mbox{g}\right)\, \right]\;\;\;.
 \ea
 Insertion of (\ref{35} - \ref{37}) into the fourth term will make it look:
 \ba
  -\,\frac{C}{L}\,\frac{1}{\sin\,I_f}\,\left(\Phi_{h_f}
 + \Phi_{\phi_f}\;\cos I_f\right)\;=\;\frac{C}{L}\;\frac{\mu_3}{\sin I_f}
 \ea
 Hence, we have for $\;I_r^{^{(rel)}}\,$:
 \ba
 \nonumber
 I_r^{^{(rel)}}=\;I+J\left(1\,-\,\frac{C}{2A}\,-\,\frac{C}{2B}\right)\,\left[\,
 \cos \mbox{g}\,-\,{\it e}\,\cos \left(2\,{\it
 l}+\mbox{g}\right)\, \right]\,+\,\frac{C}{L}\,\frac{\mu_3}{\sin I_f}
 \;-\;\alpha\;\cot I_f
 \ea
 \ba
 +\;O\left(J^2\right)\,+\,O(J\Phi/\omega)\,+\,
 O\left(\,(\Phi/\omega)^2\,\right)~~~~~
 \label{A17}
 \label{995}
 \ea
 where the parameter $\;e\;$, given by (\ref{50}),
 is the measure of triaxiality of the rotator.

 In the precessing coordinate system obtained from the inertial one by two
 Euler rotations, as in subsection A.2.2.2, we must now substitute (\ref{988})
 for $\,\alpha\,$ and (\ref{466}) for $\,\mu_3\,$, to get:
 \ba
 \nonumber
 I_r^{^{(rel)}}=I+J\left(1\,-\,\frac{C}{2A}\,-\,\frac{C}{2B}\right)\,\left[\,
 \cos \mbox{g}\,-\,{\it e}\,\cos \left(2\,{\it l}+\mbox{g}\right)\,\right]\,
 -\;\frac{C}{L}\;\dot{\pi}_1\;\cos I_f\;\sin h ~~~~~~~~~~~~~~~~~~~~~~~~~~\\
 ~\\
 \nonumber
 ~~~~~~~~+\;\frac{C}{L}\;\dot{\Pi}_1\;\left(\;\sin \pi_1\;\cos I_f\;\cos h\;+\;
 \cos\pi_1\;\sin I_f\;\right)\;+\,O\left(J^2\right)\,+\,O(J\Phi/\omega)\,+\,
 O(\,(\Phi/\omega)^2\,)~~.~~~~~~~
 \label{}
 \ea
 In the Kinoshita precessing axes obtained from the inertial ones by three
 rotations, as in subsection A.2.2.3, we should substitute (\ref{1988}) for
 $\,\alpha\,$ and (\ref{968}) for $\,\mu_3\,$, to get:
 \ba
 \nonumber
 I_r^{^{(rel)}}=I+J\left(1\,-\,\frac{C}{2A}\,-\,\frac{C}{2B}\right)\,\left[\,
 \cos \mbox{g}\,-\,{\it e}\,\cos \left(2\,{\it l}+\mbox{g}\right)\,\right]\,
 -\;\frac{C}{L}\;\dot{\pi}_1\;\cos I_f\;\sin (h\,-\,\Pi_1)~~~~~~~~~~~~~~~~~~~\\
 \\
  \nonumber
 +\,\frac{C}{L}\,\dot{\Pi}_1\left(\sin \pi_1\,\cos I_f\,\cos
 (h-\Pi_1)+\,
 \cos\pi_1\,\sin I_f -\,\sin I_f \right)\,+\,O\left(J^2\right)\,+\,O(J\Phi/\omega)\,+\,
 O(\,(\Phi/\omega)^2\,)~~.~~
 \label{}
 \ea
 To arrive to the final expression for
 $\;I_r^{^{(rel)}}\;$, we shall make use of (\ref{27} - \ref{29}). These formulae will
 enable us to substitute, in the above expression, $\,I_f\,$ and $\,h_f\,$ with $\,I\,$
 and $\,h\,$, correspondingly. All in all, in the coordinate system as in A.2.2.2 we
 have:
 \ba
 \nonumber
 I_r^{^{(rel)}}=I+J\left(1\,-\,\frac{C}{2A}\,-\,\frac{C}{2B}\right)\,\left[\,
 \cos \mbox{g}\,-\,{\it e}\,\cos \left(2\,{\it l}+\mbox{g}\right)\,\right]\,
 -\;\frac{C}{L}\;\dot{\pi}_1\;\cos I\;\sin h ~~~~~~~~~~~~~~~~~~~~\\
 \label{996}\\
 \nonumber
 ~~~~~~~~+\;\frac{C}{L}\;\dot{\Pi}_1\;\left(\;\sin \pi_1\;\cos I\;\cos h\;+\;\cos\pi_1\;
 \sin I\;\right)\;+\,O\left(J^2\right)\,+\,O(J\Phi/\omega)\,+\,O(\,(\Phi/\omega)^2\,)~~.~~
 \ea
 In the coordinate system as in A.2.2.3, we obtain:
 \ba
 \nonumber
 I_r^{^{(rel)}}=I+J\left(1\,-\,\frac{C}{2A}\,-\,\frac{C}{2B}\right)\,\left[\,
 \cos \mbox{g}\,-\,{\it e}\,\cos \left(2\,{\it l}+\mbox{g}\right)\,\right]\,
 -\;\frac{C}{L}\;\dot{\pi}_1\;\cos I\;\sin (h\,-\,\Pi_1)~~~~~~~~~~~~~~~~~~~\\
 \label{1996}\\
  \nonumber
 +\,\frac{C}{L}\,\dot{\Pi}_1\left[\sin \pi_1\,\cos I\,\cos
 (h-\Pi_1)+\,
 \cos\pi_1\,\sin I -\,\sin I \right]\,+\,O\left(J^2\right)\,+\,O(J\Phi/\omega)\,+\,
 O(\,(\Phi/\omega)^2\,)~~.~~
 \ea
 In (\ref{996}) and (\ref{1996}), the first two terms coincide with those
 given by the second of formulae (2.6) in Kinoshita (1977). They make
 $\,I_r^{^{(inert)}}\,$, while the third term is
 $\,I_r^{^{(\Phi)}}\,$.\\

 ~\\

 \noindent
{\underline{\bf{\Large{\textbf{Appendix 4.~~~Expression for
$\;h_r^{^{(rel)}}\;$}}}}}

 ~\\

Expressions (\ref{474}) and (\ref{477}) result in
 \ba
 \omega^{^{(rel)}}\;\sin \,I_r^{^{(rel)}}\;\sin \,h_r^{^{(rel)}}\;=\;\dot{I}_f\;
 \cos \,h_f\;+\;\dot{\phi}_f\;\sin \,I_f\;\sin\,h_f\;\;\;,
 \label{A19}
 \label{489}
 \label{997}
 \ea
while (\ref{475}) and (\ref{478}) entail
 \ba
 -\;\omega^{^{(rel)}}\;\sin \,I_r^{^{(rel)}}\;\cos \,h_r^{^{(rel)}}\;=\;\dot{I}_f\;
 \sin \,h_f\;-\;\dot{\phi}_f\;\sin \,I_f\;\cos\,h_f\;\;\;.\;\;\;
 \label{A19}
 \label{490}
 \label{998}
 \ea
Multiplying the former with $\,\cos h_f\,$ and the latter with
$\,\sin h_f\,$, and then summing up the two results, we arrive at
 \ba
 \dot{I}_f\;=\;\omega^{^{(rel)}}\;\sin
 I_r^{^{(rel)}}\;\sin\left(\,h_r^{^{(rel)}}\,-\;h_f\,\right)\;\;\;.
 \label{491}
 \label{999}
 \ea
 Since the difference $\;h_r^{^{(rel)}}\,-\;h_f\;$
 is expected to be of order $\;O(J)\,+\,O(\Phi/\omega)\;$, the above formula may be
 rewritten as
 \ba
 h_r^{^{(rel)}}\,-\,h_f\,=\,\frac{\dot{I}_f}{\omega^{^{(rel)}}\,\sin I_f}\,+\,
 O(J^2)\,+\,O(\,(\Phi/\omega)^2\,)\,+\,O(\,J\,\Phi/\omega\,)\,
 \label{492}
 \ea
or, according to (\ref{986}), as
 \ba
 h_r^{^{(rel)}}\,-\,h_f\,=\,\frac{1\;+\;\alpha}{\omega^{^{(inert)}}}\,\frac{1}{\sin I_f}\,
 \left(\,\frac{\partial {I_f}}{\partial t}\,+\,
 \Phi_{I_f}\,\right)\,+\,O(J^2)\,+\,O(\,(\Phi/\omega)^2\,)\,+\,O(\,J\,\Phi/\omega\,)\,
 \label{493}
 \ea
 where $\,\alpha\,$ is of order $\,O(\Phi/\omega)\,$ and therefore
 may be dropped.
Recall that, according to (\ref{47}), the absolute value of the
angular-velocity vector can be approximated, up to $\,O(J^2)\,$,
with $\,\omega^{^{(inert)}}\,\approx\,L/C\,\approx\,G/C\,$, while
$\,\sqrt{G^2\,-\,L^2}\,$ can be expressed as
$\,\sqrt{G^2\,-\,L^2}\,=\,G\;\sin
J\,\approx\,G\,J\,\approx\,L\,J\,$. Together with approximations
$\,\phi_f =\,{\it l}\,+\,\mbox{g}\,-\,J\,\cot I\,\sin
\mbox{g}\,+\,O(J^2)\,$ and
 $\,I_f\,=\,I\,+\,J\,\cos \mbox{g}\,+\,O(J^2)\,$
, it will enable us to rewrite (\ref{25}) as
 \ba
 \frac{\partial{I_{f}}}{\partial t}\;=\;L\;J\;
 \left[ \frac{\sin {\it l}\;\cos ({\it l}\,+\,g)}{A}\;-\;
 \frac{\cos {\it l}\;\sin ({\it l}\,+\,g)}{B} \right]\;+\;O(J^2)\;\;\;,~~~~~~~~~~~~~~
 \label{A33}
 \label{494}
 \ea
 insertion whereof in (\ref{494}) will then entail
 \ba
 \nonumber
 h_r^{^{(rel)}}\,-\,h_f\;=~~~~~~~~~~~~~~~~~~~~~~~~~~~~~~
 ~~~~~~~~~~~~~~~~~~~~~~~~~~~~~~~~~~~~~~~~~~~~~~~~~~~
 ~~~~~~~~~~~~~~~~~~~~~~~~~~\\
 \nonumber\\
 \nonumber\\
 \frac{J\,C}{\sin \,I_f}\,\left[ \frac{\sin {\it l}\,
 \cos ({\it l}+g)}{A}\,-\, \frac{\cos {\it l}\,\sin ({\it l}+g)}{B} \right]
 +\,\frac{C}{L}\;\frac{1}{\sin I_f}\;\Phi_{I_f}+
 O(J^2)+O(\,(\Phi/\omega)^2\,)+O(\,J\,\Phi/\omega\,)~~~~~~~
 \label{495}
 \ea
This, along with $\,h_f=\,h\,+\,J\,\sin g/\sin \,I \,+\,O(J^2)\,$
and $\,I_f\,=\,I\,+\,J\,\cos \mbox{g}\,+\,O(J^2)\,$, yields:
 \ba
 \nonumber
 h_r^{^{(rel)}}\,=\,h\,+\,J\;\frac{\sin \mbox{g}}{\sin \,I}\;+\;
 \frac{J\,C}{\sin\,
 I}\,\left\{\,\frac{\sin \it l}{A}\,\cos\left({\it
l}\,+\,\mbox{g} \right) \,-\,\frac{\cos \it l}{B}\,\sin\left({\it
l}\,+\,\mbox{g} \right)
 \,\right\}\,+\;\frac{C}{L}\;\frac{1}{\sin I}\;\Phi_{I_f}\,+\,
 O(\,.\,.\,.\,)
 \ea
 ~\\
 \ba
 \nonumber
 =\;h\;+\;\frac{J}{\sin \,I}\left\{\;\sin
 {\mbox{g}}\;+\;\frac{C}{A}\,\;\frac{\sin\left(2{\it l}\,+\,\mbox{g}\right)
 -\sin{\mbox{g}}}{2}\;-\;\frac{C}{B}\,\;\frac{\sin\left( 2{\it l}+
 \mbox{g}\right)\;+\;\sin{\mbox{g}}}{2}\;\right\}~~~~~~~~~~~~~~~~~~~~~~~~~~~~~~
 \ea
 \ba
 \nonumber
  ~~~~~\,~+\;\frac{C}{L}\;\frac{1}{\sin I}\;\left(\;
 -\;{\mu}_1\;\cos h_f\;-\;{\mu}_2\;\sin h_f
 \,\right)+\;O(J^2)\;+\;O(\,(\Phi/\omega)^2\,)\;+\;O(\,J\,\Phi/\omega\,)~~~~~~~~~
 \label{}
 \ea
 ~\\
 \ba
 \nonumber
  =\;h\;+\;\frac{J}{\sin \,I}\;\left(1\,-\,\frac{C}{2A}\,-\,\frac{C}{2B}\right)
 \;\left[\;\sin \mbox{g}\;-\;{\it e}\;\sin \left(2\;{\it l}\;+\;\mbox{g}
 \right) \;\right]~
 ~~~~~~~~~~~~~~~~~~~~~~~~~~~~~~~~~~~~~~~~~~~~~~~~~~~~~~~~~~~~~~~~~~~~~~~~~~~
 \ea
 \ba
 \nonumber
  ~~~~~\,~+\;\frac{C}{L}\;\frac{1}{\sin I}\;\left(\;
 -\;{\mu}_1\;\cos h_f\;-\;{\mu}_2\;\sin h_f
 \,\right)+\;O(J^2)\;+\;O(\,(\Phi/\omega)^2\,)\;+\;O(\,J\,\Phi/\omega\,)~~~.~~~~~
 \label{496}
 \ea
 To get the answer in the precessing coordinate axes obtained from the
 inertial ones by two Euler rotations, as in A.2.2.2, we
 substitute (\ref{466}) for $\,\mu_1\,$ and $\,\mu_2\,$. It yields:
 \ba
 h_r^{^{(rel)}}\,=\,h\;+\;\frac{J}{\sin \,I}\;\left(1\,-\,\frac{C}{2A}\,-\,
 \frac{C}{2B}\right)\;\left[\;\sin \mbox{g}\;-\;{\it e}\;\sin \left(2\;{\it l}
 \;+\;\mbox{g}\right) \;\right]~
 ~~~~~~~~~~~~~~~~~~~~~~~~~~~~~~~~~~~~~~~~~~~~~~~~~~~~~~~~~~~~~~~~~~~~~~~~~~~
 \ea
 \ba
 ~~~~~\,~-\;\dot{\pi}_1\;\frac{C}{L}\;\frac{\cos h}{\sin I}\;
 -\;\dot{\Pi}_1\;\frac{C}{L}\;\frac{\sin\pi_1\;\sin h}{\sin I}
 \;+\;O(J^2)\;+\;O(\,(\Phi/\omega)^2\,)\;+\;O(\,J\,\Phi/\omega\,)~~~.~~~~~
 \label{496}
 \ea
 To obtain the answer in the Kinoshita precessing coordinate system obtained
 from the inertial one by three Euler rotations, as in A.2.2.3, we should
 substitute (\ref{968}) for $\,\mu_1\,$ and $\,\mu_2\,$. This will entail:
\ba
 h_r^{^{(rel)}}\,=\,h\;+\;\frac{J}{\sin \,I}\;\left(1\,-\,\frac{C}{2A}\,-\,\frac{C}{2B}\right)
 \;\left[\;\sin \mbox{g}\;-\;{\it e}\;\sin \left(2\;{\it l}\;+\;\mbox{g}
 \right) \;\right]~
 ~~~~~~~~~~~~~~~~~~~~~~~~~~~~~~~~~~~~~~~~~~~~~~~~~~~~~~~~~~~~~~~~~~~~~~~~~~~
 \ea
 \ba
 ~~~~~\,~-\;\dot{\pi}_1\;\frac{C}{L}\;\frac{\cos (h-\Pi_1)}{\sin I}\,
 -\,\dot{\Pi}_1\,\frac{C}{L}\;\frac{\sin\pi_1\;\sin (h-\Pi_1)}{\sin I}
 \,+\,O(J^2)\,+\,O(\,(\Phi/\omega)^2\,)\,+\,O(\,J\,\Phi/\omega\,)~~~,~~~~
 \label{1496}
 \ea
 the triaxiality parameter $\;e\;$ being rendered by (\ref{50}). In
 (\ref{496}) and (\ref{1496}), the first two terms coincide with those
 given by the first expression of (2.6) in Kinoshita (1977). They
 constitute $\;h_r^{^{(inert)}}\;$, while the third term is
 $\;h_r^{^{\Phi}}\;$.

\end{document}